\documentclass[12pt]{iopart}

\usepackage{pstricks,graphicx,bbm,amssymb, iopams, arydshln}
\usepackage[utf8]{inputenc}
\usepackage{hyperref}
\usepackage[english]{babel}
\usepackage{color}
\usepackage{cite}

\def\Id{{\mathbbm 1}}
\def\dag{^{\dagger}}
\def\p{{\rm p}}
\def\q{{\rm q}}
\def\caseI{{\rm I}}
\def\caseII{{\rm II}}
\def\a{{\rm a}}
\def\b{{\rm b}}
\def\no{{\rm n}}
\def\unc{u}
\def\comp{c}

\def\Ttot{T_\no}
\def\Ntot{N_\no}
\def\bmsigma{\boldsymbol\sigma}
\def\sigmamA{\boldsymbol\sigma_{\rm het}}
\def\sigmamBq{\boldsymbol\sigma_{\rm a}}
\def\sigmamBp{\boldsymbol\sigma_{\rm b}}

\def\sigmaz{\boldsymbol\sigma_z}

\def\V{a}
\def\W{b}
\def\Z{z}

\begin{document}
\title[Continuous-variable quantum key distribution over multispan links$\ldots$]{Continuous-variable quantum key distribution over multispan links employing phase-insensitive and phase-sensitive amplifiers}

\author{M.~N.~Notarnicola$^{1}$, F. Cieciuch$^{2}$ and M.~Jarzyna$^{3}$}
\address{$^{1}$Dipartimento di Fisica ``Aldo Pontremoli'',
Universit\`a degli Studi di Milano and INFN Sezione di Milano, via Celoria 16, I-20133 Milano, Italy}
\address{$^{2}$Faculty of Physics, University of Warsaw, Pasteura 5, 02-093 Warszawa, Poland}
\address{$^{3}$Centre for Quantum Optical Technologies, Centre of New Technologies, University of Warsaw, Banacha 2c, 02-097 Warszawa, Poland}
\ead{\mailto{michele.notarnicola@unimi.it}, \mailto{m.jarzyna@cent.uw.edu.pl}}

\begin{abstract}
Transmission losses through optical fibers are one of the main obstacles preventing both long-distance quantum communications and continuous-variable quantum key distribution. Optical amplification provides a tool to obtain, at least partially, signal restoration.
In this work, we address a key distribution protocol over a multispan link employing either phase-insensitive or phase-sensitive amplifiers, considering Gaussian modulation of coherent states followed by homodyne detection at the receiver's side.
We perform the security analysis under both unconditional and composable security frameworks by assuming in the latter case only a single span of the whole communication link to be untrusted.
We compare the resulting key generation rate for both kinds of amplified links with the no-amplifier protocol, identifying the enhancement introduced by optical amplification.
\end{abstract}


\maketitle

\section{Introduction}
Thanks to the quantum key distribution (QKD) \cite{Gisin2002, Scarani2009, Pirandola2020} it is possible to distill a random secure key between a sender (Alice) and a receiver (Bob), communicating via an untrusted quantum channel under the control of an eavesdropper (Eve). Generally speaking, QKD protocols may be divided into two branches: discrete-variable QKD, in which qubit states are exchanged \cite{Bennett1984, Ekert1991}, and continuous-variable QKD (CV-QKD), exploiting the quadratures of a quantum optical field \cite{Grosshans2002, Laudenbach2018}. In particular, CV-QKD provides a powerful resource as it exploits coherent states of radiation and quadrature measurements \cite{Olivares2021}, thus being compatible with modulation and detection systems already employed in standard fiber-optic communications \cite{Agrawal2002}.

The milestone of CV-QKD is represented by the GG02 protocol, originally proposed by Grosshans and Grangier \cite{Grosshans2002, Laudenbach2018, Grosshans2003-1, Grosshans2003-2, Grosshans2005}, in which Alice generates coherent states by sampling a Gaussian distribution and Bob randomly implements a homodyne detection of one of the two orthogonal quadratures of the field. Later, a no-switching scheme has also been proposed in which the single quadrature measurement is replaced by double homodyne detection \cite{Weedbrook2004}.
The GG02 protocol has been widely analyzed in the {\em unconditional security} framework \cite{Laudenbach2018}, exploiting the optimality of Gaussian attacks \cite{Navascues2006, LeverrierThesis, GarciaPatron2006}. In this approach one considers the most general attack allowed by the laws of quantum mechanics, which requires the eavesdropper to possess a quantum computer-like machinery and perform any collective unitary and detection operations on many time slots. In practice, however, one can often restrict the attacks to a reasonable smaller class, either assuming a limited power by the eavesdropper or some level of trust in the infrastructure. For example, in satellite QKD a typical attack consists of detecting some part of the signal that is not captured by Bob's telescope \cite{Pirandola2021-Free}. Similarly, for the fiber-based protocols, it is reasonable to assume that an attack is performed on just a single section of the fiber and that attacker cannot easily gain access to the whole cable. For these reasons, more recently, the interest has been directed to the {\em composable security} approach, in which every step of the protocol is associated with some error, ultimately leading to an overall $\epsilon$-security \cite{Leverrier2015}. Moreover, towards a practical description, within the framework of composable security there have also been included realistic assumptions on the feasible experimental setups, such as wiretap channels \cite{Pan2020, Banaszek2021, Notarnicola2022}, restricted eavesdropping \cite{Pan2020} and different trust levels of the setup components \cite{Pirandola2021}.

Above all, in CV-QKD a crucial factor restricting security of long-distance secure communication is provided by channel losses. Indeed, a canonical model for optical fibers is provided by the thermal-loss channel, in which the channel transmissivity is exponentially decaying with the transmission distance \cite{Lodewyck2005, Lodewyck2007, Banaszek2020}.
To compensate transmission losses and restore the signal, one can employ optical amplifiers \cite{Bachor2019, Caves1982, Notarnicola2022-OPO} and consider a multispan link, that is, a periodic array of amplifiers connected by many independent thermal-loss channels. 
So far, multispan links have been investigated with the intent of increasing channel capacity \cite{Yariv1990, Antonelli2014, Jarzyna2019, Lukanowski2023}, showing that both phase-insensitive amplifiers (PIAs) \cite{Jarzyna2019} and phase-sensitive amplifiers (PSAs) \cite{Lukanowski2023} induce an exponential enhancement of the ultimate capacity being more appreciable for short-distance communication.
In contrast, in the context of CV-QKD only heralded noiseless linear amplification \cite{Ralph2009, Blandino2012, Ghalaii2020, Notarnicola2023} and quantum repeaters \cite{Furrer2018, Pirandola2019, Dias2020} have been considered. These provide innovative solutions, but are technologically challenging and far from a direct large-scale implementation.

In this paper, we address the problem of performing CV-QKD over a multispan link. We maintain the same modulation and detection schemes of GG02, that is, a Gaussian modulation of coherent states and homodyne detection, but replace the single thermal-loss channel of the original proposals with a multispan link of $M$ spans connected via either PIAs or PSAs. In particular we compare three different cases: a PIA link with random homodyne detection of one of the two orthogonal field quadratures and a PSA link with homodyne detection of either the amplified or deamplified quadrature. We compute the key generation rate (KGR), i.e. the length of the secret key per unit time slot, in both the unconditional and the composable security approaches. In the former scenario, we prove that unconditional security is improved in particular regimes only by PSA links followed by homodyne measurement of the deamplified quadrature. On the other hand, we also address composable security under restricted eavesdropping: we assume trusted amplifiers and only a single untrusted span among the whole $M$ ones composing the link. This assumption provides a simplified picture to identify the more vulnerable points of the fiber link. We compare the three discussed cases and show that amplification is helpful if the untrusted node is placed either at the beginning or the end of the link, according to the particular employed amplifier and measured quadrature. Finally, we briefly discuss the case in which Alice and Bob achieve the ultimate capacity limits discussed in \cite{Yariv1990, Antonelli2014, Jarzyna2019, Lukanowski2023}. This provides us with the ultimate enhancement in the KGR brought by the links under investigation.

The structure of the paper is the following. In Section~\ref{sec:MultiSpan} we present the structure of the multispan links under investigation. Then, in Sections~\ref{sec:UncSec} and ~\ref{sec:CompSec} we perform the security analysis under unconditional and composable security paradigms, respectively, assuming a single untrusted span. Thereafter, in Section~\ref{sec:HolevoAlice} we consider the ultimate limits obtained when Alice and Bob achieve the quantum channel capacity. Finally, in Section~\ref{sec:Conc} we draw the final conclusions and summarize the obtained results.

\section{Multispan amplified links}\label{sec:MultiSpan}

\begin{figure}
\begin{center}
\includegraphics[width=0.85\textwidth]{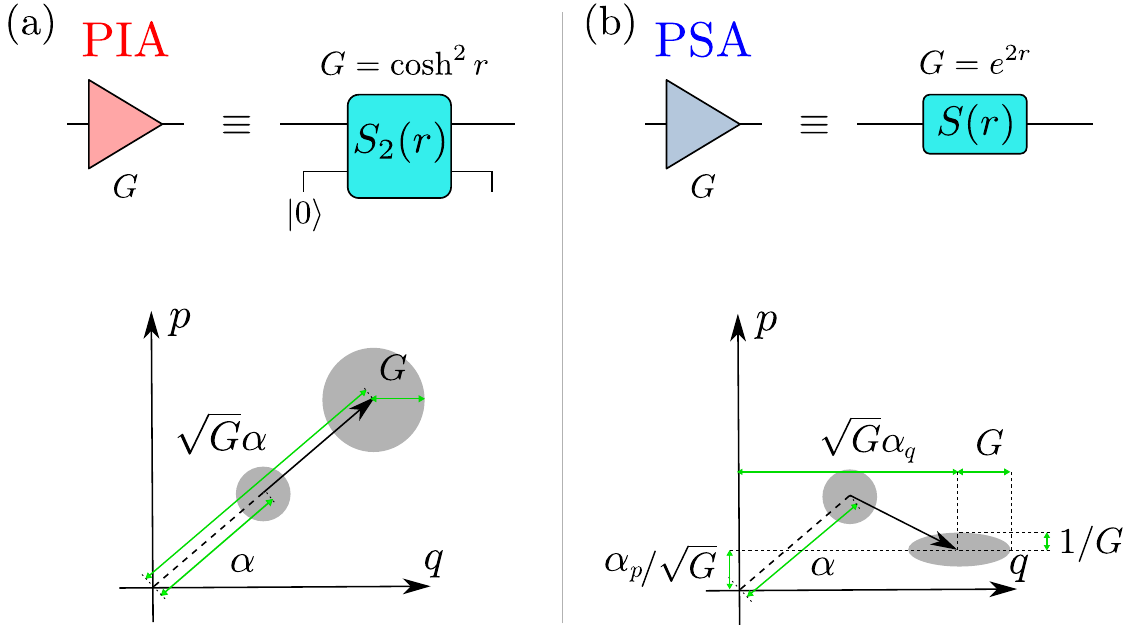}
\end{center}
\caption{Schemes of the phase-insensitive amplifier (PIA) (a) and phase-sensitive amplifier (PSA) (b) employed throughout the paper. In both the cases the amplification gain $G$ is related to the squeezing parameter $r$. The PIA applies the same amplification to both quadratures and introduces additional noise while PSA amplifies one of the quadratures and deamplifies the second rescaling the variances accordingly as seen on the example for a coherent state with amplitude $\alpha=\alpha_q+i\alpha_p$.}\label{fig:01-Amp}
\end{figure}

In this work, we address the application of multispan links employing optical amplifiers for CV-QKD. In particular, we employ the quantum amplifiers depicted in Figure~\ref{fig:01-Amp}. We consider an incoming optical mode $a$, satisfying $[a,a\dag]=1$, and its associated quadrature operators
\begin{eqnarray}
q= \sigma_0 (a+ a\dag) \quad \mbox{and} \quad p= \rmi  \sigma_0 (a\dag- a) \, ,
\end{eqnarray}
with $[q,p]=2\rmi \sigma_0^2$ and $\sigma_0^2$ being the shot-noise variance \cite{Olivares2021}. Throughout the work we will always consider shot-noise units, namely $\sigma_0^2=1$.
The goal is to amplify the mode $a$ by a gain factor $G$.
The PIA, see Figure~\ref{fig:01-Amp}(a), is implemented by coupling mode $a$ together with an ancillary mode $b$ prepared in the vacuum state $|0\rangle$ and performing a two-mode squeezing operation, namely
\begin{eqnarray}
S_2(r) = \exp\Big[ r \, \bigg(a\dag b\dag - a b\bigg) \Big] \, ,
\end{eqnarray}
$r\ge0$ being the squeezing parameter \cite{Bachor2019, Serafini2017}. The original input mode is then transformed into $a  \rightarrow \sqrt{G} \, a + \sqrt{G- 1} \,  b$, with $G=\cosh^2 r$.
Thereafter, we trace over mode $b$, ending up with an amplified signal but at the expense of introducing an ineludible added noise equal to $G-1$ to both quadratures variances.
PIAs also well describe amplification by a laser medium without optical feedback from a cavity \cite{Bachor2019, DiNicola2018}, being of particular interest for systems working at high powers, whereas their application at the quantum level is more limited due to the introduced excess noise \cite{Caves1982}.

The issue of noise may be circumvented by employing PSAs, see Figure~\ref{fig:01-Amp}(b), implemented via a unitary
single-mode squeezing operation
\begin{eqnarray}
S(r) = \exp\Bigg\{ \frac{r}{2} \, \Big[(a\dag)^2- a^2 \Big] \Bigg\} \, ,
\end{eqnarray}
$r\ge0$ \cite{Bachor2019, Serafini2017}. PSA amplifies the quadrature $q$ by a factor $\sqrt{G}=\exp(r)\ge 1$ at the expense of squeezing, i.e. deamplifying, the quadrature $p$ by $1/\sqrt{G} \le 1$. Consequently, the quadrature variances are also amplified and deamplified by $G$ and $1/G$, respectively. Crucially, the input commutation relations between the quadratures are preserved without introducing any further noise. Note also the important difference between PIA and PSA: the former is a noisy operation requiring the introduction of an additional light mode lost to the environment, which, in principle, can be intercepted by a malicious party, whereas the latter amplification scenario assumes unitary evolution which does not leak any information, thus being always trusted.

\begin{figure}
\begin{center}
\includegraphics[width=0.9\textwidth]{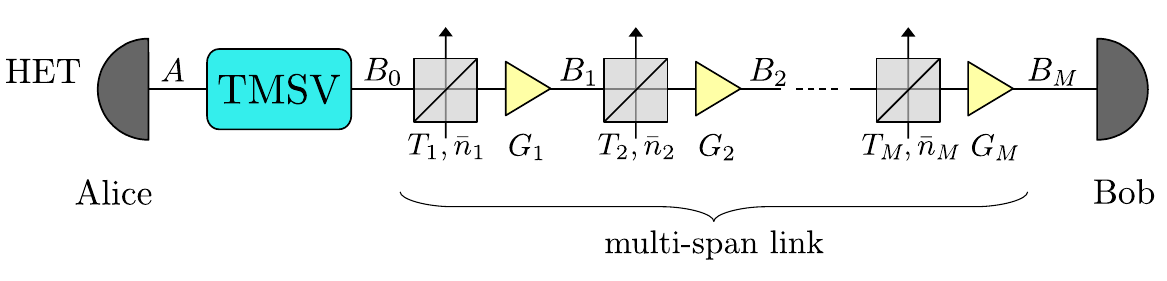}
\end{center}
\caption{Scheme of the CV-QKD protocol discussed in the present paper. A two mode squeezed vacuum state (TMSV) is distributed between Alice and Bob. Alice performs a heterodyne measurement on her mode, whereas the mode sent to Bob travels through a loss-thermal channel modeled by a series of $M$ beam splitters with transmissivities $T_j$ and added mean thermal number of photons $\bar{n}_j$. In order to counteract losses the signal after each span $B_j$ is amplified by either PIA or PSA with gain $G_j$. Finally, Bob performs a measurement which we assume to be either case $\caseI$: a random homodyne detection of quadratures $q/p$, or homodyne detection of either $q$, $\caseII\a$, or $p$, $\caseII\b$.}\label{fig:02-Proto}
\end{figure}

Given the previous considerations, in Figure~\ref{fig:02-Proto} we present the protocol discussed in the paper. We start from the GG02 scheme in its entanglement-based version \cite{Grosshans2002, Laudenbach2018, Grosshans2003-1, Grosshans2003-2, Grosshans2005}. That is, Alice has a two-mode squeezed vacuum state (TMSV) with variance $V>1$, that is,
\begin{eqnarray}\label{eq:TMSV}
|{\rm TMSV} \rangle\!\rangle =
\sqrt{1-\lambda^2}\sum_{n=0}^{\infty} \lambda^n |n\rangle |n \rangle \, ,
\end{eqnarray}
where $\lambda= \sqrt{ (V-1)/(V+1) }$ and $|n\rangle$ being the Fock state with $n$ photons \cite{Olivares2021}.
She injects the second branch into the quantum channel while performing heterodyne detection, equivalent to double homodyne, on the remaining mode, such that the conditional state sent to Bob is a coherent state. Ultimately, Bob performs a homodyne measurement on the received pulses, which in the former version of GG02 consists of a random homodyne detection of either $q$ or $p$ quadratures \cite{Grosshans2002, Grosshans2005}.

Unlike in the standard GG02 protocol, the quantum channel discussed in this work consists of a multispan link with $M$ spans alternated by optical amplifiers. 
Each span $j=1,\ldots,M$ is modeled as an independent thermal-loss channel with transmissivity $T_j\le 1$ and excess noise $\epsilon_j\ge 0$. More precisely, the optical mode entering the $j$-th link
is mixed at a beam splitter with transmissivity $T_j$ with a thermal state having $\bar{n}_j= T_j \epsilon_j/[2(1-T_j)]$ mean number of photons \cite{Olivares2021}.
Thereafter, the radiation undergoes optical amplification, either phase-insensitive or phase-sensitive,  
before being injected into the $(j+1)$-th span. 
For simplicity, here we assume both identical and equally spaced amplifiers, such that all spans have the same transmissivity $T_j=T$, added thermal noise $\bar{n}_j=\bar{n}_T$ and amplification gain $G_j=G$. Note, however, that this choice may not be the optimal arrangement \cite{Jarzyna2019}.
Then, if the total transmission distance is $L$, two neighboring amplifiers are spaced by $L/M$ and we have
\begin{eqnarray}
T= 10^{-\kappa L/(10 M)} \, ,
\end{eqnarray}
$\kappa= 0.2$ dB/km being the typical loss rate of standard optical fibers \cite{Lodewyck2005,Lodewyck2007, Banaszek2020}. Moreover, we assume the added thermal photons in each span to be equal to
 \begin{eqnarray}\label{eq:thnoise}
\bar{n}_T= \frac{T^M \epsilon}{2(1-T^M)} \, .
\end{eqnarray}
With these choices, in the absence of optical amplification, that is $G=1$, we retrieve the standard GG02 scenario, that is a single-span thermal-loss channel with total transmissivity $\Ttot= T^M$ and added noise $\Ntot=(1-\Ttot)/\Ttot+\epsilon$, $\epsilon\ge 0$ being the total excess noise \cite{Laudenbach2018, Grosshans2005}.

Starting from the scheme in Figure~\ref{fig:02-Proto}, we address three different cases, differing from one another by both the employed amplifier and the measurement implemented by Bob:
\begin{itemize}
\item Case $\caseI$: PIA link and random homodyne detection of quadratures $q/p$,
\item Case $\caseII\a$: PSA link and homodyne detection of quadrature $q$, namely the anti-squeezed quadrature,
\item Case $\caseII\b$: PSA link and homodyne detection of quadrature $p$, namely the squeezed quadrature.
\end{itemize}
Note that the presence of a PSA link makes the channel phase-sensitive, thus differentiating the behavior of quadratures $q$ and $p$. Therefore, in the presence of PSAs Bob may perform homodyne detection of a single quadrature for those experimental runs dedicated to key extraction, while homodyning both $q$ and $p$ for the channel evaluation stage, in order to fully characterize the quantum channel \cite{Laudenbach2018}.

In the following, we compute the KGR for all three cases under both unconditional and composable security scenarios.
To perform the analysis, we adopt the notation introduced in Figure~\ref{fig:02-Proto}. At first Alice has two optical modes $A$ and $B_0$ excited in the TMSV state~(\ref{eq:TMSV}). Then, the mode $B_0$ is injected into the sequence of $M$ spans. We denote by $B_j$ the optical mode coming out from the $j$-th span and subsequently amplified by the $j$-th amplifier. Finally, we refer to the last output mode as $B=B_M$.
We start by computing the mutual information shared between Alice and Bob, addressing the cases $\caseI$ and $\caseII\p$, $\p=\a,\b$, separately. The whole analysis is carried out following the Gaussian formalism, briefly reminded in~\ref{app:Gauss}.

\subsection{Case $\caseI$ : PIA link}\label{sec:MI-PIA} 
The initial state before injection into the channel is a TMSV in modes $A$ and $B_0$, completely characterized by its covariance matrix (CM)
\begin{eqnarray}
\bmsigma_{AB_0} =
\left(
\begin{array}{cc} V \, \Id_2 & Z \, \sigmaz \\ Z \, \sigmaz & V \, \Id_2 \end{array} 
\right) \, ,
\end{eqnarray}
where $Z=\sqrt{V^2-1}$, $\Id_2$ is a $2\times 2$ identity matrix and $\sigmaz$ is the Pauli $z$-matrix. 

The mode $B_0$ is injected into the noisy channel, which may be modeled via a sequence of completely positive (CP) Gaussian maps as derived in~\ref{app:CP}. Accordingly, after $M$ nodes applying PIA the state shared between Alice and Bob is still Gaussian with CM 
\begin{eqnarray}\label{eq:CM_Nnodes_PIA}
\bmsigma_{AB}^{(\caseI)}= 
\left(
\begin{array}{cc} \bmsigma_A^{(\caseI)} & \bmsigma_Z^{(\caseI)} \\[1ex] \bmsigma_Z^{(\caseI) \mathsf T} & \bmsigma_{B}^{(\caseI)} \end{array}
\right) 
=
\left(
\begin{array}{cc}
\V^{(M)} \, \Id_2 &  \Z^{(M)} \sigmaz \\
\Z^{(M)} \sigmaz & \W^{(M)} \, \Id_2  \\
\end{array}
\right)  \, ,
\end{eqnarray}
where
\numparts
\begin{eqnarray}
\V^{(M)}= V \, , \\[1ex]
\W^{(M)} = T^{(M)} \bigg[ V+ N^{(M)} \bigg]  \, , \\[1ex]
\Z^{(M)}= \sqrt{T^{(M)}} \, Z  \, , 
\end{eqnarray}
\endnumparts
and
\numparts
\begin{eqnarray}
T^{(M)}= (G T)^M \, , \\[1ex]
N^{(M)} = \frac{1}{(G T)^{M-1}} \frac{1- (GT)^M}{1-GT} \bigg[N +N_G\bigg]  \, ,
\end{eqnarray}
\endnumparts
$N= (1-T) (1+2\bar{n}_T)/T$ being the added noise introduced after the passage though a single span due to the channel thermal noise, while $N_G=(G-1)/(G T)$ is the added noise due to the PIA.
Consequently, compared to the scenario in the absence of amplifiers, the PIA link is equivalent to a thermal-loss channel with increased transmissivity $T^{(M)} \ge \Ttot$, but also increased added noise $N^{(M)}\ge \Ntot$.

After transmission, Alice performs heterodyne detection on her mode, associated with the CM $\sigmamA= \Id_2$, while Bob implements a homodyne detection of either quadrature $q$ or $p$, referred to as subcases $\a$ and $\b$, and described by the CMs
\begin{eqnarray}
\sigmamBq = \lim_{z\rightarrow 0} 
\left(
\begin{array}{cc} z & 0 \\ 0 & z^{-1}\end{array}
\right) 
\quad \mbox{and} \quad
\sigmamBp = \lim_{z\rightarrow \infty} 
\left(
\begin{array}{cc} z & 0 \\ 0 & z^{-1}\end{array}
\right) \, ,
\end{eqnarray}
respectively. Due to the symmetry of~(\ref{eq:CM_Nnodes_PIA}), the resulting statistics for both quadrtures are identical, therefore, we can safely assume that Bob always measures the quadrature $q$. In turn, the mutual information between Alice and Bob may be obtained directly from~(\ref{eq:CM_Nnodes_PIA}) as \cite{Notarnicola2023}:
\begin{eqnarray}\label{eq: IAB-PIA}
I_{AB}^{(\caseI)} (V,G) = \frac12 \log_2 \Bigg\{\frac{\det\big[\bmsigma_A^{(\caseI)}+\sigmamA\big] \det\big[\bmsigma_B^{(\caseI)}+\sigmamBq \big]}{\det\big[\bmsigma_{AB}^{(\caseI)}+(\sigmamA  \oplus \sigmamBq )\big]} \Bigg\} \, ,
\end{eqnarray}
where we highlighted the dependence on the free parameters $V$ and $G$.

\subsection{Case $\caseII$ : PSA link}\label{sec:MI-PSA} 
For cases $\caseII\p$, $\p=\a,\b$, we obtain the CM of the state shared between Alice and Bob as:
\begin{eqnarray}\label{eq:CM_Nnodes_PSA}
\bmsigma_{AB}^{(\caseII)}= 
\left(
\begin{array}{cc} \bmsigma_A^{(\caseII)} & \bmsigma_Z^{(\caseII)} \\[1ex] \bmsigma_Z^{(\caseII) \mathsf T} & \bmsigma_{B}^{(\caseII)} \end{array}
\right) 
=
\left(
\begin{array}{cccc}
\V^{(M)} &   0 & \Z_1^{(M)} & 0 \\
 0 & \V^{(M)} & 0 & -\Z_2^{(M)}  \\
\Z_1^{(M)} & 0 & \W_1^{(M)} & 0  \\
 0& -\Z_2^{(M)} & 0& \W_2^{(M)} \\
\end{array}
\right)  \, ,
\end{eqnarray}
where
\numparts
\begin{eqnarray}
\W_{1(2)}^{(M)} = T_{1(2)}^{(M)} \bigg[ V+ N_{1(2)}^{(M)} \bigg]  \, , \\[1ex]
\Z_{1(2)}^{(M)}= \sqrt{T_{1(2)}^{(M)}} \, Z  \, ,
\end{eqnarray}
\endnumparts
and
\numparts
\begin{eqnarray}
T_1^{(M)}&= (G T)^M \, , \qquad T_2^{(M)}=(G^{-1} T)^M \,,  \\[1.5ex]
N_1^{(M)} &= \frac{1}{(G T)^{M-1}} \frac{1- (GT)^M}{1-GT} N \, , \\[1.5ex]
N_2^{(M)} &= \frac{1}{(G^{-1} T)^{M-1}} \frac{1- (G^{-1}T)^M}{1-G^{-1}T} N \, ,
\end{eqnarray}
\endnumparts
with $N= (1-T) (1+2\bar{n}_T)/T$.

Unlike case $\caseI$, the PSA link is a phase-sensitive channel.
Indeed, in the presence of PSA, the quadrature $q$ exhibits an increased transmissivity $T_1^{(M)} \ge \Ttot$ and reduced added noise $N_1^{(M)} \le \Ntot$, while the quadrature $p$ shows a reduced transmissivity $T_2^{(M)} \le \Ttot$ with increased added noise $N_2^{(M)} \ge \Ntot$.
As we discuss in the following, under appropriate conditions this allows Bob to hide behind the increased noise to reduce the amount of information intercepted by an eventual eavesdropper. The mutual information for the two subcases $\p=\a,\b$ then reads:
\begin{eqnarray}\label{eq: IAB-PSA}
I_{AB}^{(\caseII\p)} (V,G) = \frac12 \log_2 \Bigg\{\frac{\det\big[\bmsigma_A^{(\caseII)}+\sigmamA\big] \det\big[\bmsigma_B^{(\caseII)}+\bmsigma_\p \big]}{\det\big[\bmsigma_{AB}^{(\caseII)}+(\sigmamA  \oplus \bmsigma_\p )\big]} \Bigg\} \, .
\end{eqnarray}

In the next sections, we will perform a security analysis of the above protocols by considering both the cases of unconditional security, where the entire channel is untrusted, and composable security, assuming that only a single span is untrusted and may be intercepted by Eve. In both scenarios, we take as a benchmark the security of the associated protocol in the absence of optical amplifiers, referred to as the ``no-amplifier protocol", in which we assume Bob to perform a random homodyne measurement of either quadrature $q$ or $p$ as in GG02.
The results of the standard no-amplifier protocol can be retrieved from both cases $\caseI$ and $\caseII$ by fixing $G=1$.

\section{Unconditional security}\label{sec:UncSec}
At first, we analyze the performance of the discussed protocol under the unconditional security approach, where the whole transmission line is supposed to be attacked by Eve. In this framework, all elements of the multispan link are assumed to be untrusted and the most powerful attack is the so-called purification attack \cite{Laudenbach2018, Grosshans2005}. That is, Eve intercepts all the lost photons and collects modes associated with the channel noise and purifies the final state shared between Alice and Bob, such that the tripartite system $ABE$ is pure \cite{Laudenbach2018}.
Under these conditions employing PIAs is useless because Eve would have access also to their purification, denoted by mode $b$ in Fig.~\ref{fig:01-Amp}(a), and extract more information with respect to the no-amplifier protocol. In contrast, case $\caseII$ is still worth of interest due to the unitarity of phase-sensitive amplification.

Considering reverse reconciliation \cite{Laudenbach2018, Grosshans2005}, for cases $\caseII\p$, $\p=\a,\b$, the KGR is given by
\begin{eqnarray}
K^{(\caseII\p)}_\unc (V,G)=\beta I_{AB}^{(\caseII\p)}(V,G)- \chi_{BE}^{(\caseII\p)}(V,G) \, ,
\end{eqnarray}
where $\beta\le 1$ is the reconciliation efficiency and $\chi_{BE}^{(\caseII\p)}(V,G)= S_E - S_{E|B}^{(\p)}$ is the Holevo information between Bob and Eve \cite{Holevo1998}, $S_E$ and $S_{E|B}^{(\p)}$ being the Von Neumann entropies of Eve's overall state and Eve's conditional state after Bob's measurement, respectively.
Due to the purification attack and the fact that Bob's measurement is represented by a $1$-rank operator, we have $S_E=S_{AB}$ and $S_{E|B}^{(\p)}=S_{A|B}^{(\p)}$, where $S_{AB}$ and $S_{A|B}^{(\p)}$ are the Von Neumann entropies of Alice and Bob's bipartite state and Alice's conditional state, respectively.
These two latter quantities can be retrieved from the CM~(\ref{eq:CM_Nnodes_PSA}), leading to:
\begin{eqnarray}\label{eq: chiBE-UNC}
\chi_{BE}^{(\caseII\p)}(V,G) = h(d_1)+ h(d_2) -h(d_3^{\, (\p)}) \, ,
\end{eqnarray}
where 
\begin{eqnarray}\label{eq:hfunc}
h(x)= \frac{x+1}{2} \log_2 \bigg( \frac{x+1}{2}\bigg) - \frac{x-1}{2} \log_2 \bigg( \frac{x-1}{2}\bigg)\, , ~
\end{eqnarray}
$d_{1}$ and $d_2$ are the symplectic eigenvalues of~(\ref{eq:CM_Nnodes_PSA}) and $d_3^{\, (\p)}= \sqrt{\det\big[ \bmsigma_{A|B}^{(\caseII\p)} \big]}$, with
\begin{eqnarray}
\bmsigma^{(\caseII\p)}_{A|B}= \bmsigma_A^{(\caseII)} - \bmsigma_Z^{(\caseII)} \Big[\bmsigma_B ^{(\caseII)}+ \bmsigma_\p \Big]^{-1} \bmsigma_Z^{(\caseII) \mathsf T} \, .
\end{eqnarray}
In particular, we have:
\begin{eqnarray}\label{eq:d3}
d_3^{\, (\a(\b))}= V \sqrt{1-\frac{Z^2}{V\bigg[V+ N_{1(2)}^{(M)}\bigg]}} \, .
\end{eqnarray}

Finally, we perform optimization over the free parameters $V$ and $G$, obtaining
\begin{eqnarray}\label{eq:KGRUnc}
K^{(\caseII\p)}_\unc = \max_{V,G} \, K^{(\caseII\p)}_\unc (V,G) \, , \qquad (\p=\a,\b) \, ,
\end{eqnarray}
subject to the set of constraints
\begin{eqnarray}\label{eq:ConditionG}
T_{1}^{(j)} \bigg[ V+ N_{1}^{(j)} \bigg] \le V \, , \quad (j=1,\ldots,M) \, ,
\end{eqnarray}
providing that throughout the channel the squeezing operation does not amplify the total optical power over its input value \cite{Jarzyna2019, Lukanowski2023}.
Since we assume all amplifiers are characterized by the same gain , it suffices to verify condition~(\ref{eq:ConditionG}) for $j=1$, satisfied if
\begin{eqnarray}\label{eq:ConditionG-PSA}
G \le G_{\rm max}^{(\caseII)} \equiv \frac{V}{1+T(V+\epsilon-1)} \, .
\end{eqnarray}

\begin{figure}
\begin{center}
\includegraphics[width=\textwidth]{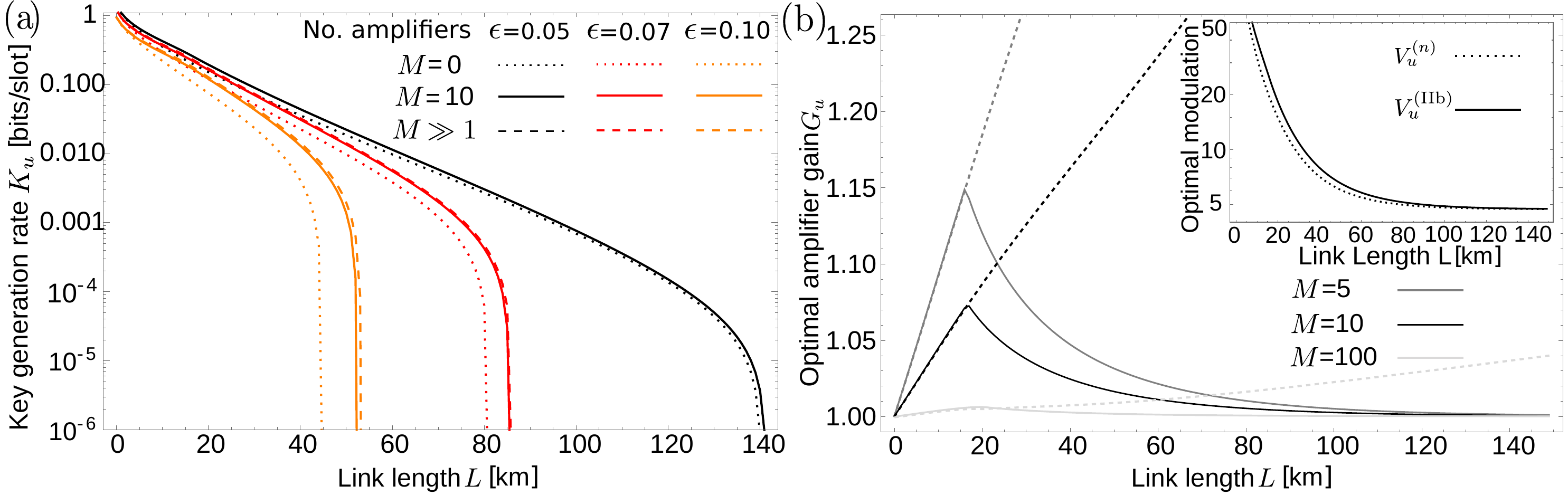}
\end{center}
\caption{(a) Optimized KGR as a function of the transmission link length $L$ for different level of external noise and number of amplifiers $M$. The case $M=0$ refers to the no-amplifier protocol. (b) Optimal amplifier gain $G^{(\caseII\b)}_\unc$ as a function of link length $L$ for different number of amplifiers $M$. The dashed lines represent the maximum attainable gain $G^{(\caseII)}_{\rm max}$ computed with the optimized modulation $V^{(\caseII\b)}_\unc$, presented in the inset. We set $\epsilon=0.05$ and $\beta=0.95$.}\label{fig:02-UncSec}
\end{figure}

In this security paradigm, the no-amplifier protocol is described by a single-span quantum channel with transmissivity $T_\no$ and added noise $N_\no$, which coincides with the GG02 protocol. The benchmark key rate $K^{(\no)}_\unc$ is obtained as:
\begin{eqnarray}
K^{(\no)}_\unc = \max_{V} \, K^{(\caseII\p)}_\unc(V,G=1) \, .
\end{eqnarray}
The obtained numerical results suggest that the optimized gain for case $\caseII\a$ is equal to $G^{(\caseII\a)}_\unc \equiv 1$ for all $L$, therefore $K^{(\caseII\a)}_\unc \equiv K^{(\no)}_\unc$ and measuring the anti-squeezed quadrature $q$ does not increase the key rate of the discussed protocol. 
On the contrary, the case $\caseII\b$ improves the security for large values of excess noise $\epsilon$, as depicted in Figure~\ref{fig:02-UncSec}(a). In this case, PSA links offer a higher KGR and, remarkably, increase the achievable maximum transmission distance, although the enhancement is relevant only for large excess noise, namely $\epsilon \gtrsim 0.05$ \cite{Roumestan2021, Roumestan2022}.

The optimized gain $G^{(\caseII\b)}_\unc$ obtained from the maximization procedure is plotted in Figure~\ref{fig:02-UncSec}(b). For small $L$, constraint~(\ref{eq:ConditionG-PSA}) leads to $G^{(\caseII\b)}_\unc= G^{(\caseII)} _{\rm max}$ and the gain increases with link length, whereas for larger $L$ it becomes a decreasing function approaching $1$ asymptotically. Moreover, $G^{(\caseII\b)}_\unc$ decreases with the number of spans $M$, as expected. Finally, the optimized modulation $V^{(\caseII\b)}_\unc$ is a decreasing function of the link length such that $V^{(\caseII\b)}_\unc \ge V^{(\no)}_\unc$, where $V^{(\no)}_\unc$ is the optimized modulation of the no-amplifier protocol.

\begin{figure}
\begin{center}
\includegraphics[width=\textwidth]{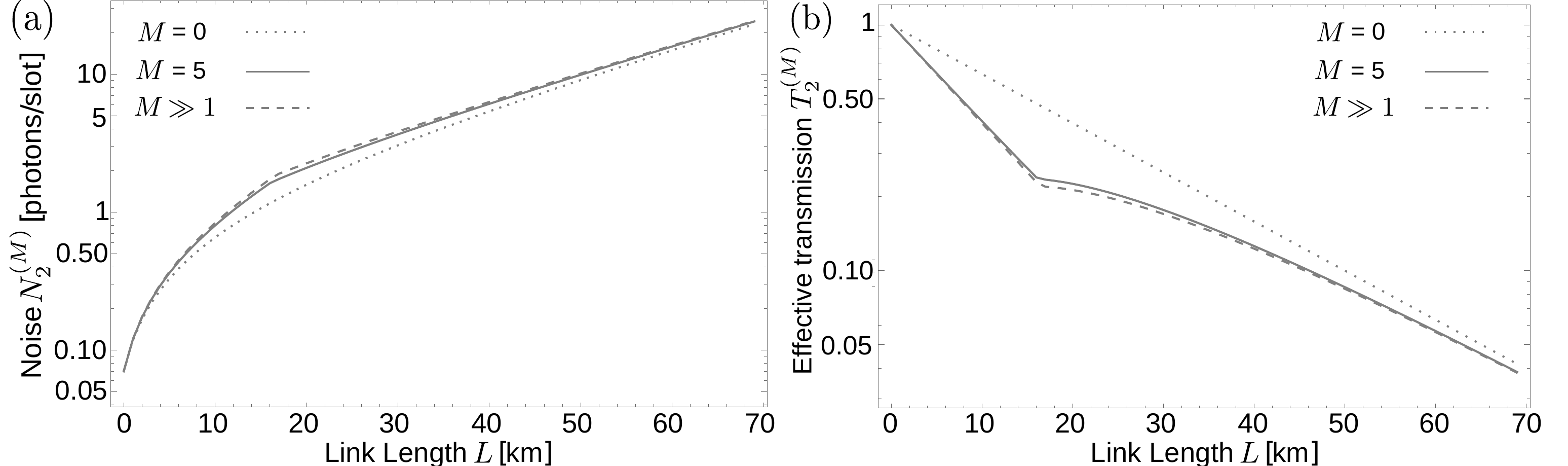}
\end{center}
\caption{Noise (a) and effective link transmission (b) as a function of link length for different number of amplifiers $M$ for $\epsilon=0.05$ and $\beta=0.95$. The case $M=0$ refers to the no-amplifier protocol.} \label{fig:03-EffPar}
\end{figure}

The physical explanation of the previous results is the following.  When measuring the squeezed quadrature $p$, Bob observes a higher added noise with respect to the standard protocol, that is, $N_2^{(M)}\ge \Ntot$, and a reduced effective transmissivity $T_2^{(M)}\le \Ttot$, as depicted in Figure~\ref{fig:03-EffPar}. In this way, he hides himself behind the noise, increasing the conditional entropy $S_{E|B}^{(\b)}$, according to~(\ref{eq:d3}), with the overall consequence of reducing the Holevo information extractable by Eve. As one may expect, the existence of the optimized gain emerges from the tradeoff between the increased noise and the reduced transmissivity, both affecting the fraction of Alice's signal actually received by Bob.

In light of this, the advantage introduced by PSAs shall increase with the number of spans $M$. In particular, we may obtain the maximum increase in KGR in the continuous-amplification limit, $M \gg 1$. Since $T^M=\Ttot$ is fixed, in this limit, up to a leading order in $M$, we have $T\approx 1$, $1-T \approx - \ln T = -(\ln \Ttot)/M$, and $G^{M}=G_\infty$. Consequently, the effective transmissivities and added noises read
\numparts
\begin{eqnarray}
T_1^{(\infty)}&= G_\infty \Ttot \,,  \qquad T_2^{(\infty)}= G^{-1}_\infty \Ttot \, , \\[1.5ex]
N_1^{(\infty)} &= \frac{1-G_\infty\Ttot}{G_\infty \Ttot} \frac{\ln \Ttot}{\ln\left(G_\infty \Ttot\right)}(1+2\bar{n}_T)\, ,  \\[1.5ex]
N_2^{(\infty)} &= \frac{1-\Ttot/G_\infty}{\Ttot/G_\infty} \frac{\ln \Ttot}{\ln\left( \Ttot/G_\infty\right)}(1+2\bar{n}_T) \, ,
\end{eqnarray}
\endnumparts
and we obtain the KGR by~(\ref{eq:KGRUnc}). These channel parameters, calculated for the resulting optimized gain $G_\infty$, are plotted in Figure~\ref{fig:03-EffPar}. Note also that even a few spans allow one to approach the continuous amplification limit.

\begin{figure}[t!]
\begin{center}
\includegraphics[width=0.6\textwidth]{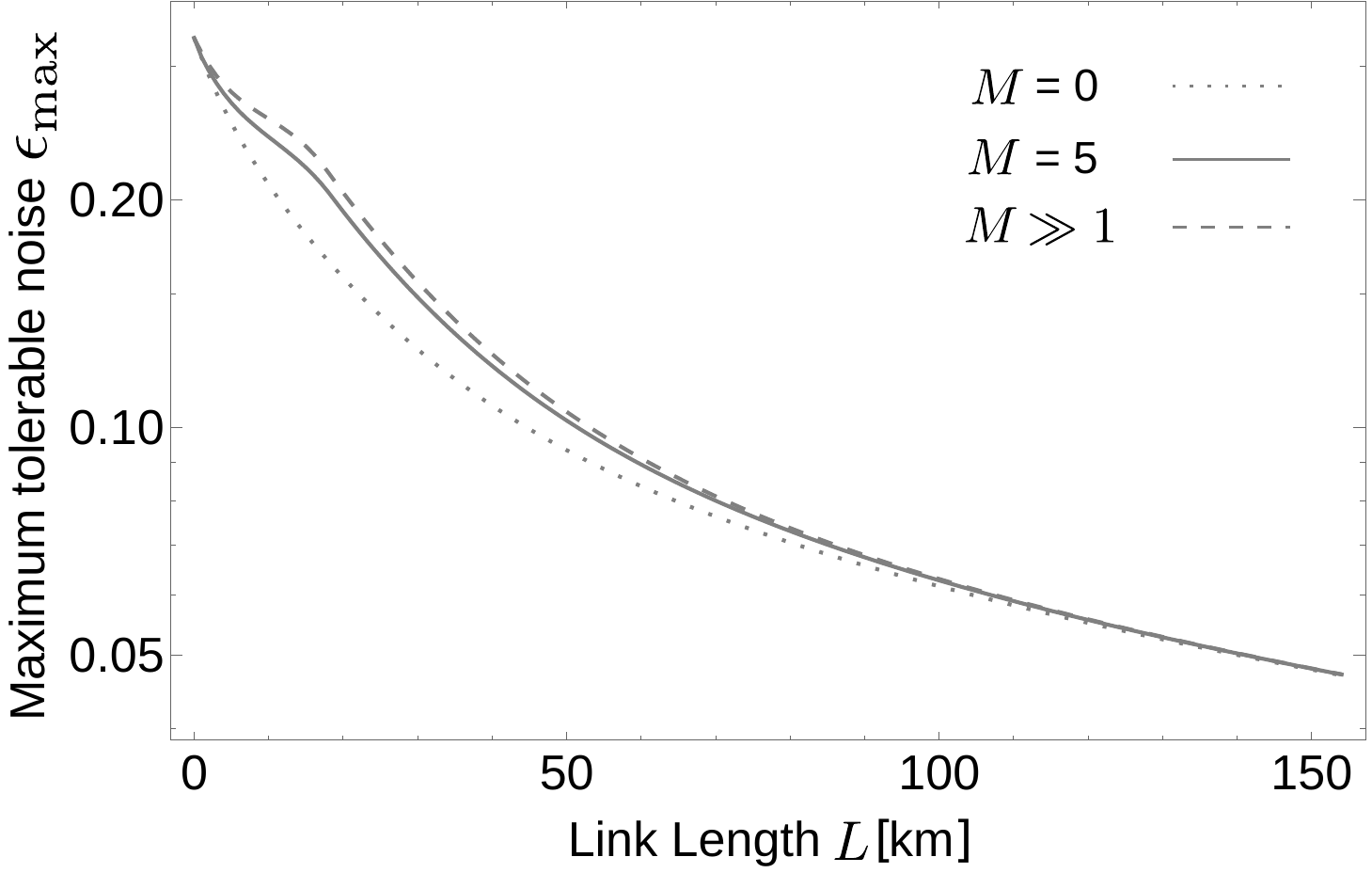}
\end{center}
\caption{Maximum tolerable noise as a function of the link length for different number of amplifiers $M$ for $\beta=0.95$. The case $M=0$ refers to the no-amplifier protocol.}\label{fig:04-MTEN}
\end{figure}

Finally, we calculate the maximum tolerable excess noise $\epsilon_{\rm max}^{(\caseII\b)}$ as a function of the transmission distance, reported in Figure~\ref{fig:04-MTEN}. It represents the maximum acceptable amount of noise to maintain a positive KGR. Consistently with the previous results, the exploitation of PSAs increases the maximum tolerable excess noise with respect to the no-amplifier scheme in the metropolitan-distance regime, as $\epsilon_{\rm max}^{(\caseII\b)} \ge \epsilon_{\rm max}^{(\no)}$. As expected, the advantage introduced increases with the number of nodes.

\section{Composable security under restricted eavesdropping: only one untrusted span}\label{sec:CompSec}

\begin{figure}[t!]
\begin{center}
\includegraphics[width=0.9\textwidth]{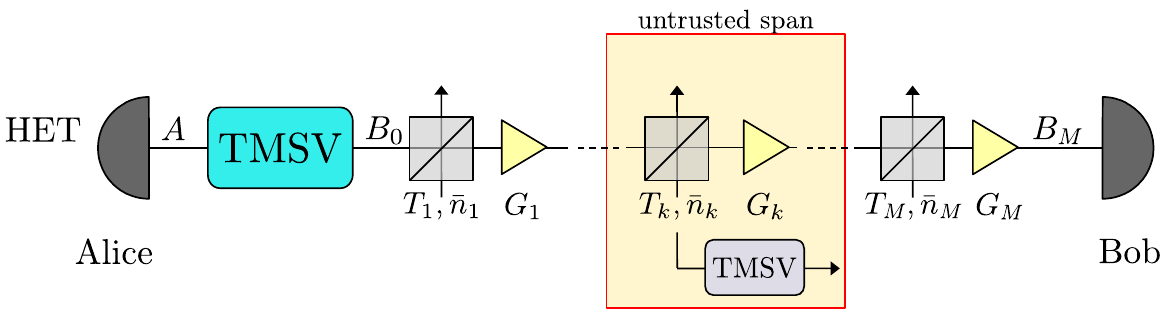}
\end{center}
\caption{Scheme of the CV-QKD protocol under restricted eavesdropping. All the amplifiers are trusted, and Eve is allowed to attack only the $k$-th span, $k=1,\ldots,M$, via active eavesdropping, that is, by injecting one arm of a TMSV state into the span, hiding herself behind the introduced excess noise. }\label{fig:06-Comp}
\end{figure}

We now discuss the second instance under investigation, the restricted eavesdropping case. In this scenario we assume Eve to attack only a single span of the link, whilst all the remaining ones as well as the employed amplifiers are considered to be trusted, thus letting our analysis to belong to the composable security framework. In turn, only a fraction $1/M$ of the whole fiber link is untrusted.
The scheme for the eavesdropping strategy under investigation is depicted in Figure~\ref{fig:06-Comp}. Across the whole channel, only the $k$ -th link, $k=1,\ldots,M$, is untrusted and may be attacked via entangling cloner attack by Eve \cite{Laudenbach2018, Pan2020}, performing active eavesdropping. That is, Eve hides herself behind the thermal noise $\bar{n}_k=\bar{n}_T$, equal to~(\ref{eq:thnoise}), by generating a TMSV state with variance $V_\epsilon= 1+2 \bar{n}_T$ in two modes $\boldsymbol{E}=(E_1,E_2)$ and injecting mode $E_1$ into the second input port of the beam splitter modeling the $k$-th span, retrieving the reflected output state. In this way she gets undetected by Alice and Bob, as performing partial trace over modes $\boldsymbol{E}$ introduces an additive thermal noise with exactly $\bar{n}_T$ mean number of photons.
In order to perform the security analysis under the above paradigm we shall compute the quantum state in Eve's possession after the entangling cloner attack. We proceed as follows, starting with the case~$\caseI$.

Since all nodes $j=1,\ldots,k-1$ are trusted, the quantum state shared by Alice and Bob injected into the $k$-th span is in the form~(\ref{eq:CM_Nnodes_PIA}), namely: 
\begin{eqnarray}
\bmsigma_{AB_{k-1}}^{(\caseI)}= 
\left(
\begin{array}{cc}
\V^{(k-1)} \, \Id_2 &  \Z^{(k-1)} \, \sigmaz \\
\Z^{(k-1)} \,\sigmaz & \W^{(k-1)} \, \Id_2  \\
\end{array}
\right) \, .
\end{eqnarray}
Instead, the CM of Eve's initial TMSV state reads:
\begin{eqnarray}
\bmsigma_{\boldsymbol{E}}= 
\left(
\begin{array}{cc}
V_\epsilon \, \Id_2 &  Z_\epsilon \, \sigmaz \\
Z_\epsilon \,\sigmaz & V_\epsilon \, \Id_2  \\
\end{array}
\right) \, ,
\end{eqnarray}
with $Z_\epsilon=\sqrt{V_\epsilon^2-1}$.
After the interference at the beam splitter, the joint quantum state of Alice, Bob and Eve is described by the CM:
\begin{eqnarray}
\bmsigma_{AB_{k} \boldsymbol{E}}^{(\caseI)} = S \, \bigg( \bmsigma_{AB_{k-1}}^{(\caseI)} \oplus \bmsigma_{\boldsymbol{E}} \bigg) \, S^{\mathsf {T}} \, ,
\end{eqnarray} 
where
\begin{eqnarray}
S=\Id_2 \oplus S_{\rm BS} \oplus \Id_2 \, ,
\end{eqnarray}
and
\begin{eqnarray}
S_{\rm BS} = \left(
\begin{array}{cc}
\sqrt{T} \, \Id_2 &  \sqrt{1-T} \, \Id_2 \\
- \sqrt{1-T} \, \Id_2 & \sqrt{T} \, \Id_2  \\
\end{array}
\right) \, 
\end{eqnarray}
is the symplectic matrix associated with the beam splitter operation \cite{Ferraro2005, Serafini2017}.
Thereafter, we let the transmitted signal pass through the remaining $M-k$ spans, applying the techniques described in~\ref{app:CP}. Ultimately, the tripartite joint state after the channel is associated with the CM:
 \begin{eqnarray}
\bmsigma_{AB \boldsymbol{E}}^{(\caseI)} = 
\left(
\begin{array}{cc}
\bmsigma_{AB}^{(\caseI)} &  \bmsigma_{C}^{(\caseI)}\\[1ex]
 \bmsigma_{C}^{(\caseI) \mathsf{T}} &\bmsigma_{\boldsymbol{E}}^{(\caseI)}  \\
\end{array}
\right) \, ,
\end{eqnarray}
with the $\bmsigma_{AB}^{(\caseI)}$ in Equation~(\ref{eq:CM_Nnodes_PIA}) and
 \begin{eqnarray}
\bmsigma_{\boldsymbol{E}}^{(\caseI)} = 
\left(
\begin{array}{cc}
\Big[ (1-T)\W^{(k-1)} + T V_\epsilon \Big]  \, \Id_2 &  \sqrt{T} Z_\epsilon \, \sigmaz \\
\sqrt{T} Z_\epsilon \, \sigmaz & V_\epsilon \, \Id_2  \\
\end{array}
\right) \, ,	\label{eq:sigmaE-PIA}
\\[2ex]
\bmsigma_{C}^{(\caseI)} = 
\left(
\begin{array}{c}
\bmsigma_{A\boldsymbol{E}}^{(\caseI)} \\[.5ex] 	\hdashline \\[-2.5ex]
\bmsigma_{B\boldsymbol{E}}^{(\caseI)}
\end{array}
\right)
= 
\left(
\begin{array}{cc}
c^{(1)} \,  \sigmaz &  \boldsymbol{0} \\[.5ex] 	\hdashline \\[-2.5ex]
c^{(2)} \, \Id_2 & c^{(3)} \,  \sigmaz  \\
\end{array}
\right) \, ,
\end{eqnarray}
being the CM of Eve's overall state and the correlation matrix between Alice and Bob and Eve, respectively, with
\numparts
\begin{eqnarray}
c^{(1)} = - \sqrt{1-T}\, \Z^{(k-1)}  \,, \\
c^{(2)} = \sqrt{(G T)^{M-k+1} (1-T)} \, \bigg[ V_\epsilon-\W^{(k-1)}\bigg] \,, \\
c^{(3)} =  \sqrt{(G T)^{M-k} G (1-T)} \,Z_\epsilon \, .
\end{eqnarray}
\endnumparts
Subsequently, after Bob's measurement Eve is left with the conditional state associated with:
\begin{eqnarray}\label{eq:sigmaEcond-PIA}
\bmsigma^{(\caseI)}_{\boldsymbol{E}|B}= \bmsigma_{\boldsymbol{E}}^{(\caseI)} - \bmsigma_{B \boldsymbol{E}}^{(\caseI) \mathsf{T}} \, \Big[\bmsigma_B ^{(\caseI)}+ \bmsigma_\a \Big]^{-1} \bmsigma_{B \boldsymbol{E}}^{(\caseI)} \, .
\end{eqnarray}

Similarly as in the unconditional security case, the KGR resulting from the present composable security analysis is given by the difference between the appropriately rescaled Alice and Bob's mutual information $I_{AB}^{(\caseI)}(V,G)$ and the Holevo information between Eve and Bob $\chi_{BE}^{(\caseI)}(V,G)$:
\begin{eqnarray}
K^{(\caseI)}_\comp (V,G)=\beta I_{AB}^{(\caseI)}(V,G)- \chi_{BE}^{(\caseI)}(V,G) \, ,
\end{eqnarray}
where $\beta$ denotes the reconciliation efficiency.  Holevo information can be written as
\begin{eqnarray}\label{eq:chiBE-comp}
\chi_{BE}^{(\caseI)}(V,G)= S_E^{(\caseI)} - S_{E|B}^{(\caseI)} = h(d_1^{\,(\caseI)})+ h(d_2^{\,(\caseI)}) -h(d_3^{(\caseI)})  -h(d_4^{(\caseI)}) \, ,
\end{eqnarray}
where $h(x)$ is the function in~(\ref{eq:hfunc}) and $d_{1(2)}^{\,(\caseI)}$ and $d_{3(4)}^{\,(\caseI)}$ are symplectic eigenvalues of the CMs~(\ref{eq:sigmaE-PIA}) and~(\ref{eq:sigmaEcond-PIA}), respectively. The resulting optimized KGR is equal to:
\begin{eqnarray}
K^{(\caseI)}_\comp = \max_{V,G} \, K^{(\caseI)}_\comp (V,G) \, ,
\end{eqnarray}
subject to the constraints $T^{(j)} [ V+ N^{(j)} ] \le V$ for all $j=1,\ldots, M$, or, equivalently,
\begin{eqnarray}\label{eq:ConditionG-PIA}
G \le G_{\rm max}^{(\caseI)} \equiv \frac{1+V}{2+T(V+\epsilon-1)} \, .
\end{eqnarray}

The same procedure may be followed to derive the key rate of case $\caseII$, identifying the corresponding CMs $\bmsigma_{\boldsymbol{E}}^{(\caseII)}$ and $\bmsigma^{(\caseII\p)}_{\boldsymbol{E}|B}$, the latter depending on the particular quadrature measured by Bob. The resulting expressions are long, and we only report them in~\ref{app:CompCM}. The corresponding KGR can be written as:
\begin{eqnarray}
K^{(\caseII\p)}_\comp (V,G)=\beta I_{AB}^{(\caseII\p)}(V,G)- \chi_{BE}^{(\caseII\p)}(V,G) \, , \quad (\p=\a,\b)\, ,
\end{eqnarray}
with the mutual information $I_{AB}^{(\caseII\p)}(V,G)$ given in~(\ref{eq: IAB-PSA}) and the Holevo information equal to
\begin{eqnarray}\label{eq:Holevo_BE_PSA}
\chi_{BE}^{(\caseII\p)}(V,G)&= S_E^{(\caseII)} - S_{E|B}^{(\caseII\p)} \nonumber \\
&= h(d_1^{\,(\caseII)})+ h(d_2^{\,(\caseII)}) -h(d_3^{\,(\caseII\p)})  -h(d_4^{\,(\caseII\p)}) \, ,
\end{eqnarray}
$d_{1(2)}^{\,(\caseII)}$ and $d_{3(4)}^{\,(\caseII\p)}$ being the symplectic eigenvalues of $\bmsigma_{\boldsymbol{E}}^{(\caseII)}$ and $\bmsigma^{(\caseII\p)}_{\boldsymbol{E}|B}$, respectively. Finally, one obtains
\begin{eqnarray}
K^{(\caseII\p)}_\comp = \max_{V,G} \, K^{(\caseII\p)}_\comp (V,G) \, ,
\end{eqnarray}
subject to the constraint~(\ref{eq:ConditionG-PSA}).

Differently from Section~\ref{sec:UncSec}, in this scenario the no-amplifier protocol is equivalent to the case of a wiretap channel under restricted eavesdropping,
in which Eve has access only to a portion $1/M$ of the fiber link \cite{Pan2020}. That is, we may model the channel as an asymmetric three-span channel composed of three beam splitters with effective transmissivities $T_{l}= T^{k-1}$, $T_{k}= T$ and $T_{r}= T^{M-k}$, and thermal noise $\bar{n}_l=\bar{n}_k=\bar{n}_r=\bar{n}_T$, respectively, in which only the central span is attacked by Eve via entangling-cloner attack. The benchmark key rate $K^{(\no)}_\comp$ is then equal to:
\begin{eqnarray}
K^{(\no)}_\comp = \max_{V} \, K^{(\caseI)}_\comp (V,G=1) \, .
\end{eqnarray}

In the next subsections, we show the obtained results, by comparing directly cases $\caseI$ and $\caseII\a$, in which the amplified quadrature is probed by Bob and, thereafter, by discussing case $\caseII\b$, where Bob detects the deamplified quadrature.

\subsection{Cases $\caseI$ and $\caseII\a$ : measuring the amplified quadrature}\label{sec:AmpQuad}
\begin{figure}
\begin{center}
\includegraphics[width=\textwidth]{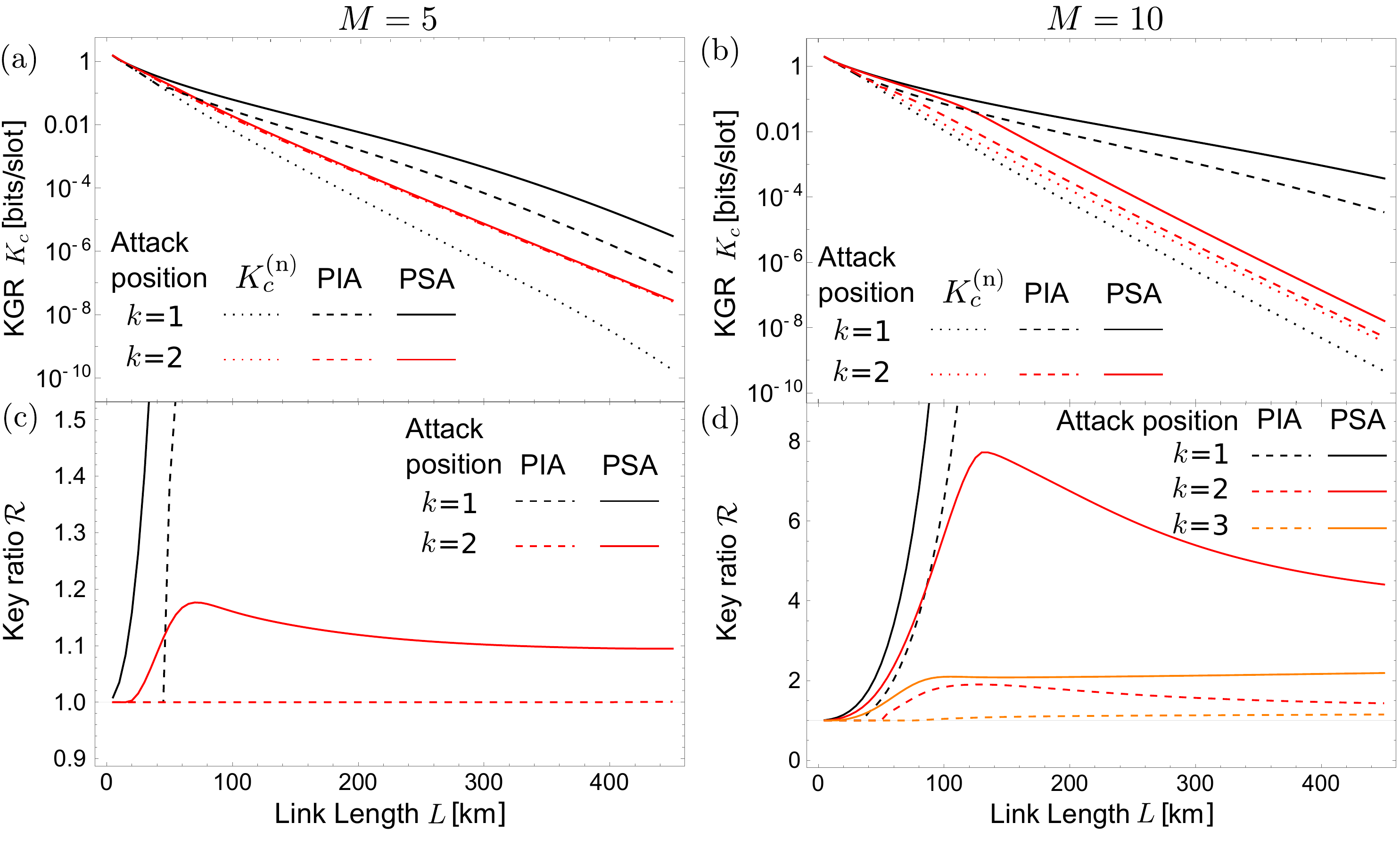}
\end{center}
\caption{Optimized KGR and key ratio for cases $\caseI$ and $\caseII$a as a function of the transmission link length $L$ for different locations of the eavesdropper for $M=5$ (a) and (c) and $M=10$ (b) and (d) respectively. We set $\epsilon=0.05$ and $\beta=0.95$.}\label{fig:07-KGRCompSec_Amp}
\end{figure}

For both the discussed cases $\caseI$ and $\caseII$, plots of the KGR $K_\comp^{(\q)}$, $\q=\caseI,\caseII\a$, are presented in Figure~\ref{fig:07-KGRCompSec_Amp} for links with $M=5$ (a) or $M=10$ (b) amplifiers and different positions $k=1,\ldots, N$ of the untrusted span, and compared to $K_\comp^{(\no)}$ for no-amplifier protocol. We underline that the results for $M=5$ and $M=10$ can be only qualitatively compared, as we keep the assumption that only one span is untrusted and, in turn, by increasing $M$ Eve becomes more and more restricted.

In general, one can observe that, when Bob measures the amplified quadrature, both PIAs and PSAs improve the KGR with respect to the no-amplifier protocol only if Eve attacks one of the first spans of the fiber link. The case $k=1$, where the first span is the untrusted one, represents the best-case scenario, where the key rate is increased by several orders of magnitude. Indeed, in this scenario the signal intercepted by Eve has not been amplified yet. Thus, Eve's overall state, described by the CM $\bmsigma_{\boldsymbol{E}}^{(\q)}$, is independent of the gain $G$ and the only effect of amplification is the reduction of the conditional entropy $S_{E|B}^{(\q)}$ appearing in the Holevo information Eqs.~(\ref{eq:chiBE-comp}) and~(\ref{eq:Holevo_BE_PSA}). On the other hand, for $k \ge 2$, amplifying Bob's received signal also increases Eve's overall entropy $S_{E}^{(\q)}$. In turn, the benefits of optical amplification are more and more reduced with increasing $k$. To better quantify this effect, we compute the ratio:
\begin{eqnarray}
{\cal R}^{(\q)} = \frac{K_\comp^{(\q)}}{K_\comp^{(\no)}} \,, \qquad (\q=\caseI,\caseII\a) \, ,
\end{eqnarray}
which is presented in Figure~\ref{fig:07-KGRCompSec_Amp}(c-d). 
All ratios are initially equal to $1$ up to a threshold distance, that is, ${\cal R}^{(\q)}=1$ if $L \le L_{\rm min}^{(\q)}$, thereafter for $k\ge 2$ they reach a maximum and then decrease towards an asymptotic value. Moreover, the key ratio ${\cal R}^{(\q)}$ decreases with increasing $k$ and there exists a threshold value $k_{\rm th}$ such that for $k \ge k_{\rm th}^{(\q)}$ we have ${\cal R}^{(\q)} \equiv 1$. Therefore, if Eve attacks a span located further, $k \ge k_{\rm th}^{(\q)}$, employing signal amplification is no longer beneficial. For link parameters values $\kappa=0.2\,\textrm{dB/km}$, $\epsilon=0.05$ and $\beta=0.95$ one obtains $k_{\rm th}^{(\caseI)}=2$ and $k_{\rm th}^{(\caseII\a)}=3$ for $M=5$, while for $M=10$ one gets $k_{\rm th}^{(\caseI)}=5$ and $k_{\rm th}^{(\caseII\a)}=8$.
Importantly, note that the performance of PIA links is always lower than PSA ones, as ${\cal R}^{(\caseI)} \le {\cal R}^{(\caseII\a)}$, $d_{\rm min}^{(\caseI)} \le d_{\rm min}^{(\caseII\a)}$ and $k_{\rm th}^{(\caseI)} \le k_{\rm th}^{(\caseII\a)}$. This is a direct consequence of the additional noise introduced by the phase-insensitive amplification process.

\begin{figure}
\begin{center}
\includegraphics[width=\textwidth]{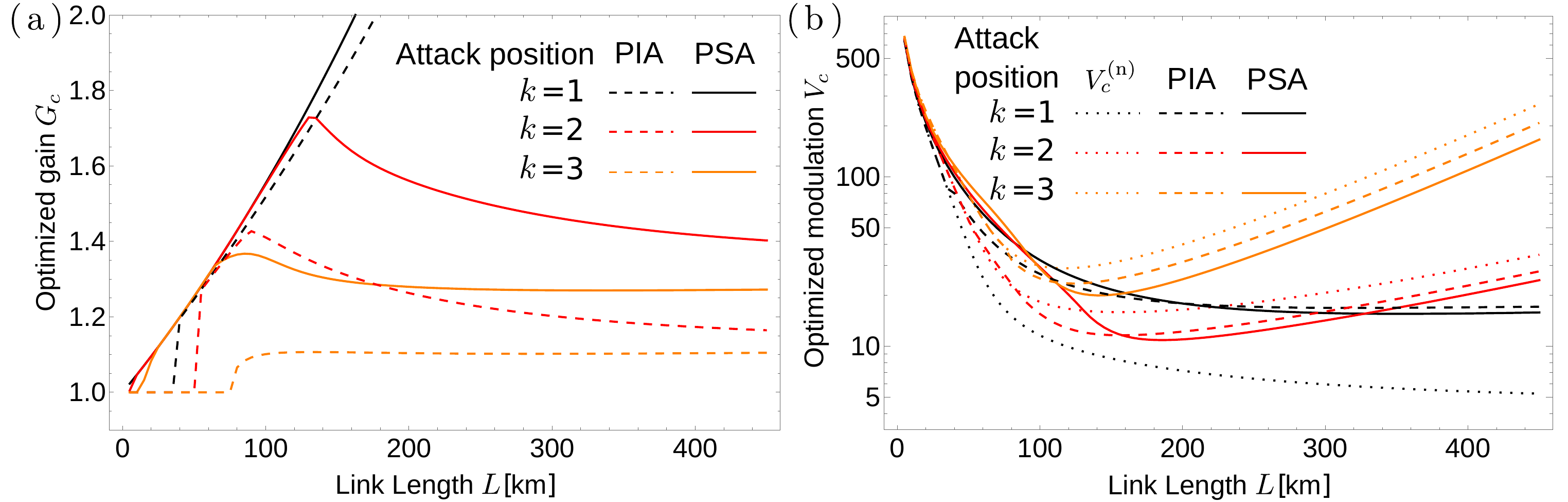}
\end{center}
\caption{Optimal amplifier gain (a) and modulation (b) for cases $\caseI$ and $\caseII$a as a function of the link length $L$ for different locations of the untrusted span $k$ for $M=10$, $\epsilon=0.05$ and $\beta=0.95$.}\label{fig:08-OptPar_CompSec_Amp}
\end{figure}

The optimized gain $G^{(\q)}_\comp$ and modulation $V^{(\q)}_\comp$, $\q=\caseI,\caseII\a$, are depicted in Figure~\ref{fig:08-OptPar_CompSec_Amp}(a) and (b), respectively.
Consistent with the results from the previous paragraph, it is optimal to not amplify the signal, i.e. $G^{(\q)}_\comp=1$, for short distances $L\le L_{\rm min}^{(\q)}$. For longer link lengths the optimal gain initially increases with $L$, following constraints~(\ref{eq:ConditionG-PSA}) and~(\ref{eq:ConditionG-PIA}), and then ultimately decreases towards an asymptotic value. The optimal gain $G^{(\q)}_\comp$ also decreases with $k$, similarly to the key ratio.
On the other hand, the behavior of optimal modulation $V^{(\q)}_\comp$ is quite peculiar. For $k=1$ it is a monotonous decreasing function of the transmission distance $L$, as obtained in Section~\ref{sec:UncSec}. The presence of optical amplifiers increases the modulation value with respect to the no-amplifier protocol, as $V^{(\q)}_\comp \ge V^{(\no)}_\comp$. On the contrary, when $k\ge2$ the situation is completely different and in the long-distance regime the optimized modulation turns out to be an increasing function of $L$. In fact, if Eve attacks one of the last spans of the communication link she intercepts a weak pulse, therefore it is possible to safely increase the input modulation variance without preventing secure communication between Alice and Bob.

When the amplified quadrature is measured, the effective transmissivity probed by Bob, namely $T^{(M)}$ and $T^{(M)}_1$ for cases $\caseI$ and $\caseII\a$ respectively, is larger with respect to the no-amplifier protocol, $T^{(M)},\,T^{(M)}_1\ge T_{\no}$. This leads to an increase of both mutual information between Alice and Bob, and, at the same time, Holevo information on Eve's side. This is because for $k\ge2$ she also receives an amplified signal. In turn, when performing optimization over the free parameters, there emerges a tradeoff between these two types of information, resulting in the key rates shown in Figure~\ref{fig:07-KGRCompSec_Amp}.
In particular, for short-distance communication, $L \le L_{\rm min}^{(\q)}$, one obtains that optical amplification is useless, $G_c^{(\q)}=1$.
The difference between cases $\caseI$ and $\caseII\a$ is due to the different impact of the added noise. In fact, for case $\caseII\a$ the added noise is rescaled with respect to the no-amplifier protocol, $N_1^{(M)}\le N_{\no}$, whilst for case $\caseI$ the noise is increased because of the additive contribution $N_G$ due to phase-insensitive amplification, $N^{(M)}\ge N_{\no}$. In the latter case the (incoherent) added contribution $N_G$ detriments the mutual information between Alice and Bob, being less than its counterpart of case $\caseII\a$. Ultimately, this leads to a reduced performance of PIA links with respect to PSA  ones. 

\subsection{Case $\caseII\b$ : measuring the deamplified quadrature}\label{sec:deAmpQuad}

\begin{figure}
\begin{center}
\includegraphics[width=\textwidth]{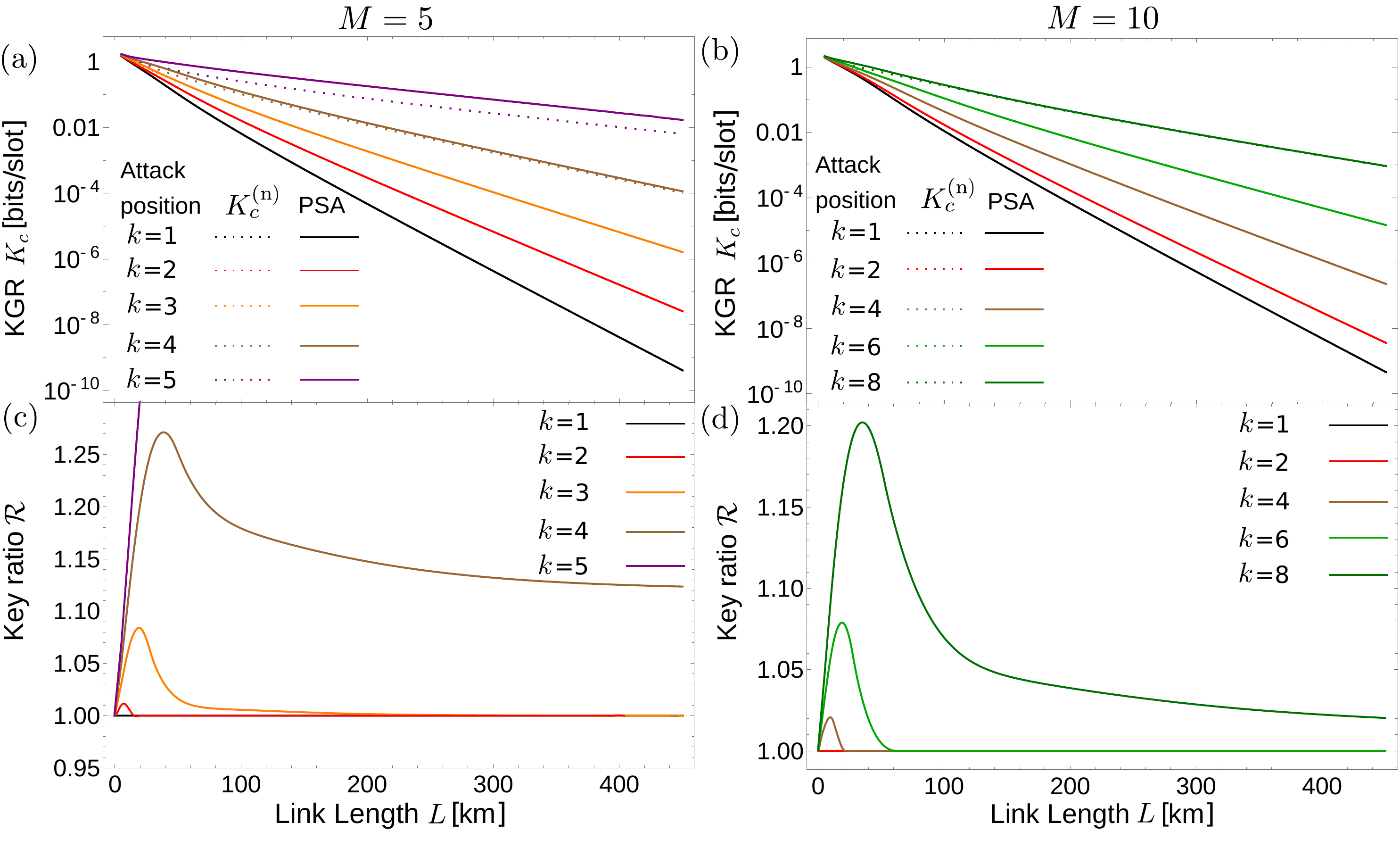}
\end{center}
\caption{Optimized KGR and key ratio for case $\caseII$b as a function of the transmission link length $L$ for different locations of the eavesdropper for $M=5$ (a) and (c) and $M=10$ (b) and (d) respectively. We set $\epsilon=0.05$ and $\beta=0.95$.}\label{fig:09-KGRCompSec_deAmp}
\end{figure}

The KGR $K_\comp^{(\caseII\b)}$ for the Bob's measurement of the deamplified quadrature, $\caseII\b$,  is depicted in Figure~\ref{fig:09-KGRCompSec_deAmp} for links with $M=5$ (a) or $M=10$ (b) amplifiers and different positions $k=1,\ldots, M$ of the untrusted span, together with the key ratio
\begin{eqnarray}
{\cal R}^{(\caseII\b)} = \frac{K_\comp^{(\caseII\b)}}{K_\comp^{(\no)}} \,.
\end{eqnarray}
The scenario is reversed with respect to the previous section. Indeed, when Bob probes the squeezed (i.e. deamplified) quadrature, PSA links improve the resulting KGR if Eve attacks one of the last spans of the channel. The best-case scenario is provided by $k=M$, in which the KGR increases by more than an order of magnitude. Consequently, and in contrast to the results from Section~\ref{sec:AmpQuad}, one observes enhancement in the key ratio ${\cal R}^{(\caseII\b)}$ with increasing $k$.
In this scenario the PSA becomes useless if Eve attacks the first span for all $M$, namely ${\cal R}^{(\caseII\b)} \equiv 1$, since in this case she intercepts the pulse before all amplifiers and therefore, deamplifying the signal only reduces the mutual information between Alice and Bob, maintaining a higher Holevo information at Eve's side.
On the other hand, for $k \ge 2$, deamplifying Bob's signal also reduces Eve's extracted information, thus leading to ${\cal R}^{(\caseII\b)} \ge 1$. 
In particular, there exists a threshold attack location $k_{\rm th}^{(\caseII\b)}$ such that for $k \le k_{\rm th}^{(\caseII\b)}$ one has ${\cal R}^{(\caseII\b)} \equiv 1$, being equal to $k_{\rm th}^{(\caseII\b)}= 1$ for $M=5$ and $k_{\rm th}^{(\caseII\b)}=2$  for $M=10$. 
For eavesdropping performed on a span located further within the link $k \ge k_{\rm th}^{(\caseII\b)}$ all key ratios exhibit a maximum and then decrease towards an asymptotic value, equal to $1$ for locations closer to the threshold value or greater than $1$ for those placed further, implying an improvement of security in the long-distance regime brought by the PSA link.

\begin{figure}
\begin{center}
\includegraphics[width=\textwidth]{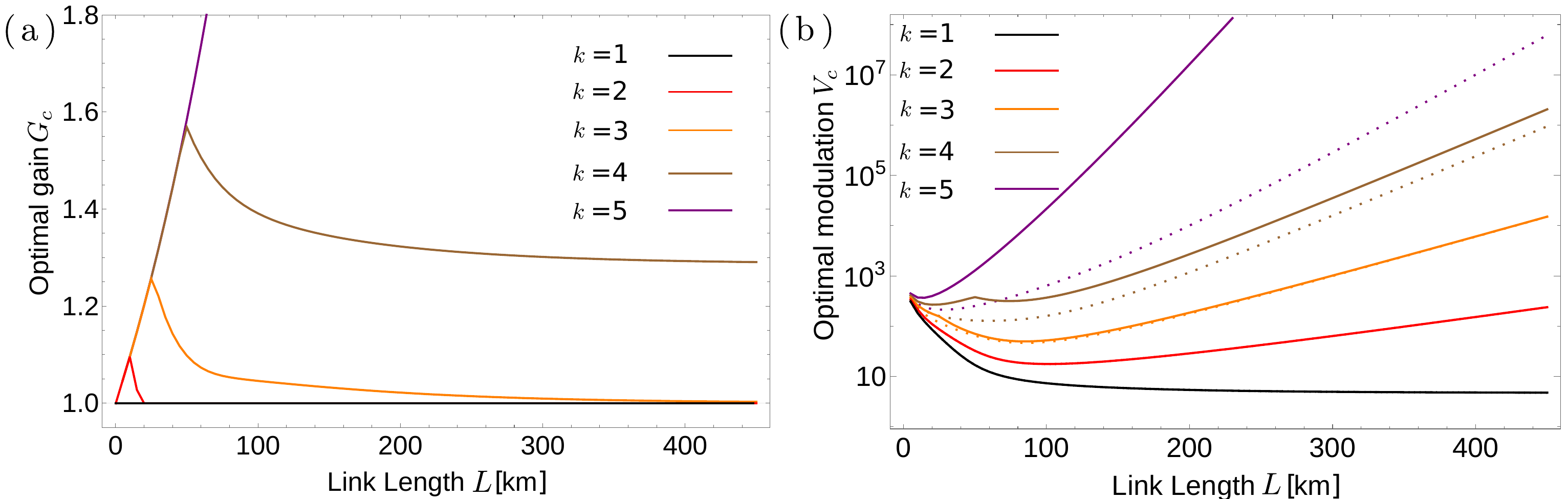}
\end{center}
\caption{Optimal amplifier gain (a) and modulation (b) for case $\caseII$b as a function of the link length $L$ for different locations of the untrusted span $k$ for $M=10$, $\epsilon=0.05$ and $\beta=0.95$.}\label{fig:10-OptPar_CompSec_deAmp}
\end{figure}

In Figure~\ref{fig:10-OptPar_CompSec_deAmp}(a) and (b), one can see the optimized gain $G^{(\caseII\b)}_\comp$ and modulation $V^{(\caseII\b)}_\comp$, respectively.
We see that amplification is not beneficial, $G^{(\caseII\b)}_\comp \equiv 1$ , for eavesdropping performed on initial spans $k\le k_{\rm th}^{(\caseII\b)}$, whereas for attacks on latter spans the optimal gain increases with the link length following constraint~(\ref{eq:ConditionG-PSA}), until finally decreasing towards an asymptotic value. In accordance with the previous results, one needs to employ the stronger optimal amplification the further the eavesdropped span is located.
The optimized modulation increases with respect to the no-amplifier protocol $V^{(\caseII\b)}_\comp \ge V^{(\no)}_\comp$. Similarly to the results obtained in Section~\ref{sec:AmpQuad}, it is a decreasing function of the link length if the attack is performed on the first span, whilst it becomes non-monotonous for $k\ge 2$, increasing in the long-distance regime.

The physical meaning of these results is analogous to these obtained in Section~\ref{sec:UncSec}. Indeed, the case~$\caseII\b$ is associated with a reduced transmissivity with respect to the no-amplifier protocol, $T^{(M)}_2\le T_{\no}$, and amplified added noise $N^{(M)}_2\ge N_{\no}$. Therefore, for $k\ge 2$ by employing PSAs Bob accepts to reduce the extracted mutual information, in order to increase the conditional entropy $S_{E|B}^{(\caseII\b)}$, resulting in a lower Holevo information between Eve and himself. The tradeoff between these two quantities is such that for $k \ge k_{\rm th}^{(\caseII\b)}$ one has $G^{(\caseII\b)}_\comp \ge 1$ and PSA links increase the obtained KGR.

\section{Ultimate key rate limits}\label{sec:HolevoAlice}

In the analysis above, we considered the entanglement-based CV-QKD protocol where Alice and Bob perform heterodyne and homodyne detection, respectively, which in the prepare-and-measure picture is equivalent to considering coherent-state encoding followed by quadrature detection \cite{Grosshans2002, Laudenbach2018, Grosshans2003-1}. In turn, the resulting mutual information depends only on the signal-to-noise ratio, according to the Shannon-Hartley theorem \cite{Essiambre2010}, as derived in~(\ref{eq: IAB-PIA}) and~(\ref{eq: IAB-PSA}).
However, from the perspective of quantum communication, we may also consider the fundamental quantum limit, namely the Holevo information between Alice and Bob \cite{Holevo1998}. This ultimate capacity provides an upper bound on the achievable mutual information which has been also investigated for multispan links employing either PIAs or PSAs \cite{Jarzyna2019, Lukanowski2023}.

We embed this approach within the CV-QKD framework by considering an equivalent protocol to that of Figures~\ref{fig:02-Proto} and~\ref{fig:06-Comp}, in which Bob still performs the homodyne measurement of quadratures $q$ and $p$, but Alice replaces her heterodyne detection with the proper measurement achieving the Holevo bound. In this way, we compute the Holevo information $\chi_{AB}$ by interpreting Bob's measurement as a state-preparation process. Ultimately, we obtain an upper bound on the KGR which allows to highlight the ultimate limits on KGR of multispan links. For the sake of simplicity, here we will discuss only the case of a PIA link under restricted eavesdropping with the same assumptions as in Section~\ref{sec:CompSec}, namely a single untrusted span.

For case $\caseI$, the ultimate KGR reads:
\begin{eqnarray}
\widetilde{K}^{(\caseI)}_\comp = \max_{V,G} \Bigg[ \beta \chi_{AB}^{(\caseI)}(V,G)- \chi_{BE}^{(\caseI)}(V,G) \Bigg] \, ,
\end{eqnarray}
with the $\chi_{BE}^{(\caseI)}$ in~(\ref{eq:chiBE-comp}) and $\chi_{AB}^{(\caseI)}$ being the Holevo capacity \cite{Jarzyna2019} retrievable from~(\ref{eq:CM_Nnodes_PIA}):
\begin{eqnarray}
\chi_{AB}^{(\caseI)}(V,G) = S_{A}^{(\caseI)} - S_{A|B}^{(\caseI)} = h(\tilde{d}_1)-h(\tilde{d}_2)\, ,
\end{eqnarray}
where $h(x)$ is the function in~(\ref{eq:hfunc}), $\tilde{d}_1= \sqrt{\det\big[ \bmsigma_{A}^{(\caseI)} \big]}= V$ and $\tilde{d}_2= \sqrt{\det\big[ \bmsigma_{A|B}^{(\caseI)} \big]}$, where
\begin{eqnarray}
\bmsigma^{(\caseI)}_{A|B}= \bmsigma_A^{(\caseI)} - \bmsigma_Z^{(\caseI)} \Big[\bmsigma_B ^{(\caseI)}+ \bmsigma_\a \Big]^{-1} \bmsigma_Z^{(\caseI) \mathsf T} \, .
\end{eqnarray}

The maximization procedure is, once again, subject to constraint~(\ref{eq:ConditionG-PIA}).

\begin{figure}
\begin{center}
\includegraphics[width=\textwidth]{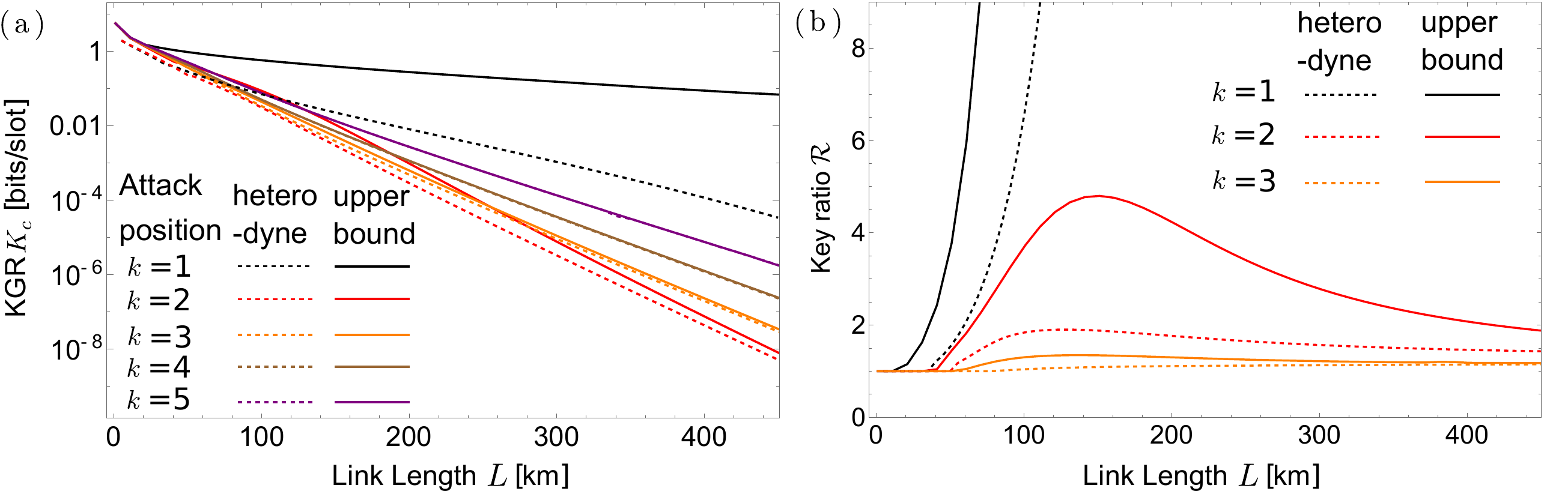}
\end{center}
\caption{Upper bound on KGR (a) and corresponding key ratio (b) for PIA as a function of link length for $M=10$ nodes, $\epsilon=0.05$ and $\beta=0.95$.}\label{fig:11-HolevoKGR}
\end{figure}

It is seen in Fig.~\ref{fig:11-HolevoKGR}(a) that there is a considerable gap between the upper bound $\widetilde{K}^{(\caseI)}_\comp$ and the performance of heterodyne receiver $K^{(\caseI)}_\comp$ for few attacks performed on few initial nodes. For further spans, the difference disappears, as indicated by the key ratios ${\cal R}$ with respect to the associated no-amplifier protocol, depicted in Fig.~\ref{fig:11-HolevoKGR}(b). This means that, even in the best theoretical scenario, the advantage originating from PIA vanishes for attacks on further parts of the link. A possible remedy may be to employ PSA but obtaining the ultimate bound in such a scenario requires considerable effort since the resulting quantum channel is phase sensitive \cite{Lukanowski2023, Schafer2016}.

\section{Conclusions}\label{sec:Conc}
In this paper we have addressed the exploitation of multispan amplified links for CV-QKD. In particular, we have discussed three cases: PIA link with random homodyne detection of $q/p$ quadratures and PSA link with measurement of either quadrature $q$ or $p$. In the unconditional security approach we showed that the KGR is improved with respect to the standard no-amplifier protocol only for the scenario in which Bob measures the deamplified quadrature. This enhancement is noticeable especially in the presence of large excess noise, $\epsilon \gtrsim 0.05$.

We have also investigated the KGR in the composable security framework, by assuming that all amplifiers and spans except one are trusted. We showed that the position of the untrusted span greatly affects the potential enhancement offered by amplification. In particular, for the cases with PIA and random homodyne measurement and with PSA and measurement of the amplified quadrature one observes an enhancement in the KGR only if one of the first spans is attacked, namely $k\le k_{\rm th}^{(\q)}$, whereas for case with PSA and detection of the deamplified quadrature the improvement is present if the attacked span is in the latter part of the link, $k\ge k_{\rm th}^{(\caseII\b)}$.
Finally, we have addressed the case in which Alice and Bob achieve the Holevo capacity, thus providing an upper bound for the security, highlighting the ultimate enhancement that may be brought by PIA links.

The results of the paper present a detailed analysis of the improvement and the limits offered by optical amplifiers for quantum secure communications under realistic assumptions and pave the way for future developments in the framework of composable security CV-QKD. In particular, the advantage given by PSA, being a phase-sensitive operation, may be potentially further boosted by employing squeezed states \cite{Gottesman2001, Derkach2020}.

\ack
This project has received funding from the European Union’s Horizon Europe research and innovation programme under the project “Quantum Security Networks Partnership” (QSNP, grant agreement No 101114043).

\appendix
\section{Brief review of the Gaussian formalism}\label{app:Gauss}
As discussed in the main text, we exploit the Gaussian formalism to perform the security analysis \cite{Serafini2017, Ferraro2005}. Here, we present the main tools to retrieve the obtained results.
We consider a $n$-mode bosonic system, described by bosonic annihilation operators $a_k$ satisfying the canonical commutation relations $[a_k,a_l]=0$, $[a_k,a_l\dag]=\delta_{kl}$, and by the quadrature operators
\begin{eqnarray}
    q_k= \sigma_0 (a_k+ a_k\dag) \quad \mbox{and} \quad p_k= \rmi \sigma_0 (a_k\dag-a_k) \, ,
\end{eqnarray}
such that $[q_k,p_l]=2 \rmi \sigma_0^2 \delta_{kl}$, where we adopt shot-noise units, $\sigma_0^2=1$. A more compact notation is obtained by introducing the vector operator $\hat{\mathbf{r}}= (q_1,p_1,q_2,p_2,...,q_n,p_n)^\mathsf{T}$. 

A Gaussian state $\rho_G$ is a quantum state associated with a Gaussian Wigner function
\begin{eqnarray}
    W[\rho_G](\mathbf{r}) =  \frac{1}{(2\pi)^n \sqrt{\det(\boldsymbol{\sigma})}} \,  \exp \bigg[ 
    - \displaystyle \frac12 (\mathbf{r}-\mathbf{R})^\mathsf{T} \, \boldsymbol{\sigma}^{-1} \, (\mathbf{r}-\mathbf{R})
    \bigg] \, ,
\end{eqnarray}
where $\mathbf{r}^\mathsf{T}= (x_1,y_1,x_2,y_2,...,x_n,y_n) \in {\mathbbm R}^{2n}$, and
\begin{eqnarray}
    \mathbf{R}= \Tr[\rho_G \, \hat{\mathbf{r}}]
\end{eqnarray}
is the first moment vector and
\begin{eqnarray}
    \boldsymbol{\sigma} = \frac12 \Tr \bigg[ \rho_G \, \{(\hat{\mathbf{r}}-\mathbf{X}),(\hat{\mathbf{r}}-\mathbf{X})^\mathsf{T}\} \bigg] \, 
\end{eqnarray}
is the $2n\times 2n$ covariance matrix (CM), where $\{A,B\}=AB+BA$ is the anti-commutator of $A$ and $B$.
Thus, a Gaussian state is completely characterized by its prime moments and its covariance matrix.

Gaussian dynamics, i.e. unitary evolution generated by bilinear Hamiltonians, is associated with a symplectic matrix $S$ such that if the input state is Gaussian the evolved state is still Gaussian with \cite{Serafini2017,Ferraro2005}:
\begin{eqnarray}
    \mathbf{R} \rightarrow S \, \mathbf{R} \, \quad \mbox{and} \quad
    \boldsymbol\sigma \rightarrow S \, \boldsymbol\sigma \, S^\mathsf{T} \, . 
\end{eqnarray}
On the contrary, Gaussian completely-positive (CP) maps are associated with a pair of matrices $X$ and $Y$ such that the evolved state is characterized by:
\begin{eqnarray}
    \mathbf{R} \rightarrow X \, \mathbf{R} \, \quad \mbox{and} \quad
    \boldsymbol\sigma \rightarrow X \, \boldsymbol\sigma \, X^\mathsf{T} + Y \, , 
\end{eqnarray}
where $Y +\rmi \Omega \ge \rmi X\Omega X^{\mathsf{T}}$, $\Omega$ being the $2n\times 2n$ symplectic form \cite{Serafini2017,Ferraro2005}.

Finally, we discuss the case of conditional dynamics. In the paper, we consider a bipartite system $AB$, where subsystems $A$ and $B$ are composed of $n_{A}$ an $n_B$ modes, respectively. We consider a Gaussian state $\rho_{AB}$ with prime moments $\mathbf{R}=(\mathbf{R}_A,\mathbf{R}_B)$ and CM (written in block form)
\begin{eqnarray}
\boldsymbol{\sigma} = 
\left(
\begin{array}{cc} 
\boldsymbol{\sigma}_A & \boldsymbol{\sigma}_{C} \\ 
\boldsymbol{\sigma}_{C}^\mathsf{T} & \boldsymbol{\sigma}_B 
\end{array} 
\right)\, .
\end{eqnarray}
We now perform a Gaussian measurement on subsystem $B$ associated with the CM $\boldsymbol{\sigma}_m$, obtaining outcome $\mathbf{r}_m \in \mathbb{R}^{2n_B}$. Then, the conditional state $\rho_{A|\mathbf{r}_m}$ on mode $A$ is still a Gaussian state with CM $\boldsymbol\sigma_{A|\mathbf{r}_m}$ and first moment vector $\mathbf{R}_{A|\mathbf{r}_m}$ given by:
\begin{eqnarray}
    \boldsymbol{\sigma}_{A|\mathbf{r}_m} = \boldsymbol{\sigma}_A - \boldsymbol{\sigma}_{C} (\boldsymbol{\sigma}_B+\boldsymbol{\sigma}_m)^{-1} \boldsymbol{\sigma}_{C}^\mathsf{T}\,,
\end{eqnarray}
and
\begin{eqnarray}
    \mathbf{R}_{A|\mathbf{r}_m} = \mathbf{R}_A + \boldsymbol{\sigma}_{AB} (\boldsymbol{\sigma}_B+\boldsymbol{\sigma}_m)^{-1}(\mathbf{r}_m-\mathbf{R}_B)\,,
\end{eqnarray}
respectively \cite{Serafini2017,Ferraro2005}.

\section{Derivation of the Gaussian CP map for the multispan link}\label{app:CP}
In the paper we discuss CV-QKD over multispan links composed of either PIAs or PSAs connected via a sequence of thermal-loss (TL) channels. In the following we report the structure of the quantum CP maps associated with each of these components.
 
A thermal-loss channel with transmissivity $T \le1$ and thermal noise $\bar{n}_T$ is described via a Gaussian CP map associated with the matrices \cite{Serafini2017}:
\begin{eqnarray}
X_{\rm TL} = \sqrt{T} \, \Id_2 \quad \mbox{and} \quad Y_{\rm TL} = (1-T)(1+2\bar{n}_T) \Id_2 \,,
\end{eqnarray}  
$\Id_2$ being the $2\times 2$ identity matrix.

As regards optical amplification, PIA are described by the Gaussian CP map \cite{Serafini2017}:
\begin{eqnarray}
X_{\rm PIA} = \sqrt{G} \, \Id_2 \quad \mbox{and} \quad Y_{\rm PIA} = (G-1) \Id_2 \,,
\end{eqnarray}  
$G\ge 1$ being the amplification gain, whilst PSA are unitary maps, thus completely described by the symplectic matrix \cite{Ferraro2005}:
\begin{eqnarray}
S_{\rm PSA}= 
\left(
\begin{array}{cc} 
G & 0 \\ 
0 & G^{-1} 
\end{array} 
\right)\, .
\end{eqnarray}

We remark that in the main text we consider a two-mode state $AB$, where only branch $B$ undergoes a Gaussian evolution. Thus, according to the Gaussian formalism, the bipartite map will be associated with the matrices $X_{AB}= \Id_2 \oplus X_{\p}$ and $Y_{AB}=  \boldsymbol{0} \oplus Y_{\p}$ for $\p= {\rm TL, PIA}$, and $S_{AB}=\Id_2 \oplus S_{\rm PSA}$ for the phase-sensitive amplification map.

\section{Composable security analysis for case $\caseII$}\label{app:CompCM}

The key rate for cases $\caseII\p$, $\p=\a,\b$, may be computed by following the same procedure described in Section~\ref{sec:CompSec}, leading to the joint state of the three parties:
 \begin{eqnarray}
\bmsigma_{AB \boldsymbol{E}}^{(\caseII)} = 
\left(
\begin{array}{cc}
\bmsigma_{AB}^{(\caseII)} &  \bmsigma_{C}^{(\caseII)}\\[1ex]
 \bmsigma_{C}^{(\caseII) \mathsf{T}} &\bmsigma_{\boldsymbol{E}}^{(\caseII)}  \\
\end{array}
\right) \, ,
\end{eqnarray}
with the $\bmsigma_{AB}^{(\caseII)}$ in Equation~(\ref{eq:CM_Nnodes_PSA}) and
 \begin{eqnarray}
\bmsigma_{\boldsymbol{E}}^{(\caseII)} = 
\left(
\begin{array}{cccc}
e_1 & 0 &  \sqrt{T} Z_\epsilon & 0 \\
0 & e_2 &  0 & -\sqrt{T} Z_\epsilon \\
\sqrt{T} Z_\epsilon & 0 & V_\epsilon & 0  \\
0 & -\sqrt{T} Z_\epsilon & 0 & V_\epsilon
\end{array}
\right) \, ,	\label{eq:sigmaE-PSA}
\\[2ex]
\bmsigma_{C}^{(\caseII)} = 
\left(
\begin{array}{c}
\bmsigma_{A\boldsymbol{E}}^{(\caseII)} \\[1ex] 	\hdashline \\[-2ex]
\bmsigma_{B\boldsymbol{E}}^{(\caseII)}
\end{array}
\right)
= 
\left(
\begin{array}{cccc}
c^{(1)}_1 & 0 &0 & 0 \\
0 & -c^{(1)}_2 & 0 & 0 \\[1ex] 	\hdashline \\[-2ex]
c^{(2)}_1 & 0 & c^{(3)}_1 & 0  \\
0 & c^{(2)}_2 & 0 & -c^{(3)}_2
\end{array}
\right) \, ,
\end{eqnarray}
with
\numparts
\begin{eqnarray}
e_{1(2)} = \Big[ (1-T)\W_{1(2)}^{(k-1)} + T V_\epsilon \Big] \, , \\[1ex]
c^{(1)}_{1(2)} = - \sqrt{1-T}\, \Z^{(k-1)}_{1(2)}  \,,\\[1ex]
c^{(2)}_{1} = \sqrt{(G T)^{M-k+1} (1-T)} \, \bigg[ V_\epsilon-\W^{(k-1)}_{1}\bigg] \,, \\[1ex]
c^{(2)}_{2} = \sqrt{(G^{-1} T)^{M-k+1} (1-T)} \, \bigg[ V_\epsilon-\W^{(k-1)}_{2}\bigg] \,, \\[1ex]
c^{(3)}_{1} =  \sqrt{(G T)^{M-k} G (1-T)} \,Z_\epsilon \,, \\[1ex]
c^{(3)}_{2} =  \sqrt{(G^{-1} T)^{M-k} G^{-1} (1-T)} \,Z_\epsilon \, .
\end{eqnarray}
\endnumparts
Finally, Eve's conditional CM reads:
\begin{eqnarray}\label{eq:sigmaEcond-PSA}
\bmsigma^{(\caseII\p)}_{\boldsymbol{E}|B}= \bmsigma_{\boldsymbol{E}}^{(\caseII)} - \bmsigma_{B \boldsymbol{E}}^{(\caseII) \mathsf{T}} \, \Big[\bmsigma_B ^{(\caseII)}+ \bmsigma_\p \Big]^{-1} \bmsigma_{B \boldsymbol{E}}^{(\caseII)} \, , \quad (\p=\a,\b) \, .
\end{eqnarray}

\section*{References} 

\bibliographystyle{iopart-num}

\providecommand{\newblock}{}

\end{document}